\documentclass[sigconf]{acmart}
\usepackage{amsmath,amsfonts}
\usepackage{lettrine}
\usepackage{subfigure}
\usepackage{multirow}
\usepackage{multicol}
\usepackage{tablefootnote}
\usepackage{enumerate}
\usepackage{enumitem}
\usepackage{extarrows}
\usepackage{booktabs}
\usepackage{mathtools}
\usepackage{multirow}
\usepackage{caption}
\usepackage{graphicx}
\usepackage{url} % simple URL typesetting
\usepackage{pifont}
\usepackage{extarrows}
\usepackage{amsfonts}
\usepackage{marvosym}
\AtBeginDocument{%
  \providecommand\BibTeX{{%
    \normalfont B\kern-0.5em{\scshape i\kern-0.25em b}\kern-0.8em\TeX}}}

\setcopyright{acmcopyright}
\copyrightyear{2024}
\acmYear{2024}
\acmDOI{XXXXXXX.XXXXXXX}

\acmConference[Conference WWW 2024]{The Web Conference}{May 13--17,
  2024}{Singapore}
\acmPrice{15.00}
\acmISBN{xxx-x-xxxx-XXXX-X/xx/xx}

\begin{document}

\title{Enhancing Cross-Domain Click-Through Rate Prediction via Explicit Feature Augmentation}

\author{Xu Chen, Zida Cheng, Jiangchao Yao, Chen Ju, Weilin Huang, Jinsong Lan, Xiaoyi Zeng, Shuai Xiao}
\email{{huaisong.cx,chengzida.czd,weilin.hwl,jinsonglan.ljs,yuanhan,shuai.xsh}@taobao.com,{sunarker,ju\_chen}@sjtu.edu.cn}
\authornote{corresponding author}
\affiliation{%
  \institution{Alibaba Group and Shanghai Jiao Tong University}
  % \streetaddress{P.O. Box 1212}
  \country{China}
}

\renewcommand{\shortauthors}{Xu Chen et al.}

\begin{abstract}
% Data sparsity is an important issue for click-through rate (CTR) prediction, particularly when user-item interactions is too sparse to learn a reliable model. Recently, many works on cross-domain CTR (CDCTR) prediction have been developed in an effort to leverage meaningful data from a related domain. 
Cross-domain CTR (CDCTR) prediction is an important research topic that studies how to leverage meaningful data from a related domain to help CTR prediction in target domain. 
Most existing CDCTR works design implicit ways to transfer knowledge across domains such as parameter-sharing that regularizes the model training in target domain. More effectively, recent researchers propose explicit techniques to extract user interest knowledge and transfer this knowledge to target domain. However, the proposed method mainly faces two issues: 1)
it usually requires a super domain, \textit{i.e.} an extremely large source domain, to cover most users or items of target domain, and 2) the extracted user interest knowledge is static no matter what the context is in target domain. These limitations motivate us to develop a more \emph{flexible} and \emph{efficient} technique to explicitly transfer knowledge.
In this work, we propose a cross-domain augmentation network (CDAnet) being able to perform explicit knowledge transfer between two domains. 
Specifically, CDAnet contains a designed translation network and an augmentation network which are trained sequentially. The translation network computes latent features from two domains and learns meaningful cross-domain knowledge of each input in target domain by using a designed cross-supervised feature translator. Later the augmentation network employs the explicit cross-domain knowledge as augmented information to boost the target domain CTR prediction.
Through extensive experiments on two public benchmarks and one industrial production dataset, we show CDAnet can learn meaningful translated features and largely improve the performance of CTR prediction. CDAnet has been conducted online A/B test in image2product retrieval at Taobao app, bringing an absolute \textbf{0.11 point} CTR improvement, a relative \textbf{0.64\%} deal growth and a relative \textbf{1.26\%} GMV increase. 
% It has also been successfully deployed online, serving billions of consumers.

% However, most existing CDCTR works have an impractical limitation that requires homogeneous inputs (\textit{i.e.} shared feature fields) across domains, and CDCTR with heterogeneous inputs (\textit{i.e.} varying feature fields) across domains has not been widely explored but is an urgent and important research problem.
% In this work, we propose a cross-domain augmentation network (CDAnet) being able to perform knowledge transfer between two domains with \textit{heterogeneous inputs}. 
% Specifically, CDAnet contains a designed translation network and an augmentation network which are trained sequentially.
% The translation network is able to compute features from two domains with heterogeneous inputs separately by designing two independent branches, 
% and then learn meaningful cross-domain knowledge using a designed cross-supervised feature translator. Later the augmentation network encodes the learned cross-domain knowledge via feature translation performed in the latent space and fine-tune the model for final CTR prediction.
% Through extensive experiments on two public benchmarks and one industrial production dataset, we show CDAnet can learn meaningful translated features and largely improve the performance of CTR prediction. CDAnet has been conducted online A/B test in image2product retrieval at Taobao app over 20 days, bringing an absolute \textbf{0.11 point} CTR improvement, a relative \textbf{0.64\%} deal growth and a relative \textbf{1.26\%} GMV increase.
\end{abstract}






% \received{12 November 2023}
\received[accepted]{1 February 2024}

\maketitle

\section{Introduction}
Click-through rate (CTR) prediction which estimates the probability of a user clicking on a candidate item, has a vital role in online services like recommendation, retrieval, and advertising~\cite{cheng2016wide,zhou2018deep,ouyang2019deep}. 
% For example, as shown in Figure~\ref{figure:taobao_app_example}, Taobao, as one of the largest e-commercial platforms in the world, has a list of application domains such as text2product retrieval and image2product retrieval. Each domain has its own CTR prediction model, which is termed as single-domain CTR prediction. 
It has been pointed out that data sparsity is a key issue that significantly limits the improvement of single-domain CTR models~\cite{li2015click}, and recent effort has been devoted to improving the target domain CTR models by leveraging data from other related domains~\cite{li2015click,ouyang2020minet,liu2022continual,zhang2022keep,sheng2021one}. 
% \begin{figure}[t]
% %第一行 loss
% \centering
% \begin{minipage}[t]{0.22\textwidth}
% \centering
% \includegraphics[width=0.85\textwidth]{images/text2product_search.pdf}
% % \vspace{-10pt}
% \caption*{(a) text2product retrieval}
% \end{minipage}
% % \hspace{5pt}
% \begin{minipage}[t]{0.22\textwidth}
% \centering
% \includegraphics[width=0.85\textwidth]{images/image2product_search.pdf}
% % \vspace{-10pt}
% \caption*{(b) image2product retrieval}
% \end{minipage}
% % \vspace{-8pt}
% \caption{An example of text2product retrieval and image2product retrieval at Taobao app.
% \label{figure:taobao_app_example}
% % \vspace{-12pt}
% }
% \end{figure}

%Recent works~\cite{li2015click,ouyang2020minet,liu2022continual,zhang2022keep,sheng2021one} point out that the data sparsity issue is a key obstacle that hinders the improvement of single-domain CTR prediction~\cite{li2015click} and propose to leverage other related domain's data to enhance the CTR prediction in target domain.

Various cross-domain CTR prediction (CDCTR) methods have recently been developed, which can be roughly categorized into two groups: \textit{joint training} and \textit{pre-training \& fine-tuning}. 
The \textit{joint training} approach is developed by combining multiple CTR objectives from different domains into a single optimization process. It usually has shared parameters to build connections, and transfer the learned knowledge across different domains like MiNet~\cite{ouyang2020minet}, DDTCDR~\cite{li2019ddtcdr} and STAR~\cite{sheng2021one}. 
However, since two domains usually have different objectives in optimization, these methods usually suffer from an optimization conflict problem, which might lead to a negative transfer result~\cite{sener2018multi,vandenhende2021multi}. To deal with this issue, a number of recent approaches study more advanced mixture-of-expert~\cite{masoudnia2014mixture,yuksel2012twenty} parameter-sharing technique for jointly optimizing CDCTR objectives~\cite{ma2018modeling,tang2020progressive,tang2020progressive}.
Further, researchers believe that the knowledge are encoded in the network parameters and they propose many \textit{pre-training \& fine-tuning} models for CDCTR. 
The \textit{pre-training \& fine-tuning} method~\cite{liu2022continual,zhang2022keep} often has two stages, by training a CTR model sequentially in a source domain and then in a target domain, where the performance in the target domain can be generally improved by leveraging model parameters pre-trained from the source domain. This two-stage learning style enables the model to have only one objective at each training stage, which significantly reduces the impact of negative transfer.  It is also parameter-efficient and has been widely applied at Taobao platform.
Especially for KEEP~\cite{zhang2022keep}, it first extracts  user interest knowledge from super domain which is an extremely large domain that covers most users or items in target domain. Then, with the knowledge extraction module frozen, it feeds the knowledge to target click prediction model by a plug-in network. KEEP provides a good way to utilize the huge knowledge in super-domain and has achieved remarkable performance in industrial applications.
% it consists of a supervised pre-training knowledge extraction module and a plug-in network that incorporates the extracted knowledge. At the first stage, knowledge about user interest representation is extracted from super-domain in a supervised pre-training manner. Afterward, at the second stage, with the knowledge extraction module frozen, such knowledge is fused with the target click prediction model by using a plug-in network. 
% KEEP provides a good way to utilize the huge knowledge in super-domain and benefit the downstream target domain. 
% In \textit{pre-training \& fine-tuning}, only one objective is utilized for optimization at each training stage, which significantly reduces the impact of negative transfer. The \textit{pre-training \& fine-tuning} style is also parameter-efficient and has been widely applied at Taobao platform.

%other popular method is \textit{pre-training \& fine-tuning} which trains the target domain model with pre-trained source domain model parameters. 
%In either pre-training or fine-tuning stage, only one objective is utilized for optimization and the impact of negative transfer can be largely reduced. Accordingly, this method has been widely applied at Taobao platform~\cite{liu2022continual,zhang2022keep}.

Importantly, most of the above methods actually design different implicit ways to transfer knowledge. The employed parameter-sharing technique in both \textit{joint training} and \textit{pre-training \& fine-tuning} can regularize the target model's parameter learning, and enable the target model to have less probability of over-fitting suspicious features. Recent researchers designed explicit ways to more efficiently transfer knowledge.
% A more efficient way is to design techniques to explicitly transfer knowledge. 
For example, KEEP~\cite{zhang2022keep} explicitly extracts user interests and feeds these interests into the downstream target model for CTR prediction. However, it still has two limitations. 1) \textbf{Super-domain data requirement}: The model requires a super-domain that covers most users or items over the target domain so that the extracted user interests could be statistically useful in the target model learning, which is not always easily met in many business cases. 
2) \textbf{Static user interest}: The extracted user interest keeps unchanged for the target domain model even when the user is in different contexts (\textit{e.g.} item features, page statistic features). While static user interests under different contexts could result in inferior performance and context-aware user interest enables the model to better capture different importance of input features~\cite{hou2023deep,wang2022enhancing}.
These motivates us to design a more \emph{flexible} and \emph{efficient} method to make explicit knowledge transfer for CDCTR.
In this work, we propose novel cross-domain augmentation networks (CDAnet),
% consisting of a translation network and an augmentation network, 
which explicitly transfers knowledge from source domain to target domain. CDAnet is performed sequentially by first explicitly learning cross-domain knowledge and then encoding the learned knowledge into the target model via cross-domain augmentation. 
Specifically, a translation network is designed to encode the inputs (including user features, item features, etc.) from different domains, and learn feature translation in latent space via a designed cross-supervised feature translator.
Then cross-domain augmentation is performed in the augmentation network by augmenting target domain samples in latent space with their additional translated latent features. \emph{This explicitly encodes the knowledge learned from the source domain, providing diverse yet meaningful additional view information of each input in target domain and improving the fine-tuning on target model.} When performing translation learning, CDAnet has no 
overlap requirement on users or items between domains, and thus has \textbf{no super-domain data requirement}, providing more \emph{flexible} usage. Moreover, the inputs including various features of target domain are together translated in translation network, and thus user interest is \textbf{context-aware} in target model fine-tuning, exerting more \emph{efficient} knowledge transfer.
% via the pre-trained translation network. 
% Then cross-domain feature augmentation is performed in the fine-tuning stage by augmenting target domain samples in latent space via the pre-trained translation network. 
% This implicitly encodes the knowledge learned from the source domain, providing diverse yet meaningful additional information for improving the fine-tuning on the target model. 
% The details of our CDAnet are presented in Fig.X   
%
%In this paper, we propose a novel translation then augmentation network (CDAnet) to encourage efficient knowledge transfer even if with heterogeneous input features across domains, and benefit the target domain CTR prediction.
%Specifically, CDAnet is a two-stage learning process including a translation stage and an augmentation stage. In translation stage, the translation network learns how to translate the latent features from target domain to source domain by a cross-supervised translator.
%Then in augmentation stage, the augmentation network utilizes the pre-trained translation network parameters, together with the additional translated latent features to augment the target domain CTR training. 
% A learning comparison between CDAnet and other methods is shown in Figure~\ref{figure:general_comparison}.
Through experiments, we demonstrate that CDAnet can largely improve the performance of CTR prediction, and it has been conducted online A/B test in image2product retrieval at Taobao app, bringing an absolute \textbf{0.11 point} CTR improvement, with a relative \textbf{0.64\%} deal growth and a relative \textbf{1.26\%} GMV increase. It has been successfully deployed online, serving hundreds of millions of consumers. 
In a nutshell, the contributions of this work are summarized as follows:
\begin{itemize}
    % \vspace{-6pt}
    \item We advocate to enhance CDCTR with explicit feature augmentation and propose CDAnet consisting of a designed translation network and an augmentation network. CDAnet explicitly translates the input of target domain into source domain as additional view information, and employ this information as augmented features to boost the target domain model training. 
    % To the best of our knowledge, CDAnet is the first work that introduces the feature translation idea to CDCTR topic.
    \item Extensive experiments on various datasets demonstrate the effectiveness of the proposed method. Our CDAnet has been deployed in image2product retrieval at Taobao app, and achieved obvious improvements on CTR, deal and GMV. Through empirical studies, we show that CDAnet is able to learn meaningful translated features for boosted CTR improvement.
    % Further, by transferring the rich and huge data knowledge from text2product domain to our image2product domain,we found it can even improve the data self-circulation problem for our image2product retrieval system.
    % \vspace{-6pt}
\end{itemize}

%\hfill Eindhoven
 
%\hfill March 4, 2021

\section{Related Work}
% \subsection{Single-domain CTR}
\textbf{Single-domain CTR}: Click-through rate (CTR) prediction plays a vital role in various online services, such as modern search engines, recommendation systems and online advertising. Previous works usually combine logistic regression~\cite{richardson2007predicting} and feature engineering for CTR prediction. These methods often lack the ability to model sophisticated feature interactions, and heavily rely on human labor of designing features.
With the excellent feature learning ability of deep neural networks (DNN), deep learning approaches have been extensively studied on CTR prediction, and recent works focus on applying DNN for learning feature interactions~\cite{cheng2016wide,10.5555/3172077.3172127,wang2017deep}.
For example, Cheng \textit{et al.}~\cite{cheng2016wide} combined  shallow linear models and deep non-linear networks to capture both low and high-order features, while the power of factorization machine~\cite{rendle2010factorization} and deep networks are combined for CTR prediction in~\cite{10.5555/3172077.3172127}. 
Wang \textit{et al.}~\cite{wang2017deep} designed a deep \& cross network to learn bounded-degree feature interactions. 
Further, some works study how to model richer information for CTR tasks. For instance, DIN~\cite{zhou2018deep} and DIEN~\cite{zhou2019deep} were proposed to capture user interests based on historical click behaviors, and DSTN~\cite{ouyang2019deep} takes contextual ads when modeling user behaviors.\\

% \subsection{Cross-domain CTR} 
\noindent\textbf{Cross-domain CTR}: Single-domain CTR prediction suffers from a data sparsity issue because user behaviors in a real-world system are usually extremely sparse and have single-domain data bias. Accordingly, cross-domain CTR prediction is developed to leverage user behaviors in a relevant but data-rich domain, \textit{i.e.} source domain, to facilitate learning in the target domain.

A \emph{joint learning} method was recently developed and has become a representative approach for cross-domain CTR. For example, STAR~\cite{sheng2021one} is a star topology model that contains a centered network shared by different domains, with multiple domain-specific networks tailored for each domain. 
In MiNet~\cite{ouyang2020minet}, Ouyang \textit{et al.} attempted to explore auxiliary data (\textit{e.g.} historical user behaviors and ad title) from a source domain to improve the performance of a target domain. 
Meanwhile, a number of cross-domain recommendation (CDR) methods have been developed~\cite{hu2018conet,li2019ddtcdr,yuan2019darec,chen2020towards}, which can be naturally introduced to cross-domain CTR problems. 
% For instance, a deep cross connection network is introduced in~\cite{hu2018conet}, and it is able to transfer user rating patterns across different domains. 
For instance, in DDTCDR~\cite{li2019ddtcdr}, Li \textit{et al.} proposed a deep dual transfer network that can bi-directionally transfer information across domains in an iterative style. However, when such models benefit from sharing parameters to transfer knowledge, they meanwhile face the problem of gradient interference issue~\cite{sener2018multi,wang2020gradient,yu2020gradient} because they have two different CTR objectives during joint learning. 
To handle the issue, 
MMOE~\cite{ma2018modeling} employs a more advanced shared mixture-of-expert model, which allows for automatically allocating model parameters and alleviating task conflicts in optimization.
To decouple learning task-specific and task-shared information more explicitly, PLE~\cite{tang2020progressive} separates the network of task-shared components and task-specific components, and then adopts a progressive routing mechanism being able to extract and separate deeper semantic knowledge gradually. Further, Zhang \textit{et al.}~\cite{10.1145/3580305.3599758} highlighted that different source domains should contribute differently to the target domain when transferring knowledge. Then, they proposed CCTL that could evaluate the information gain of the source domain on the target domain and adjust the information transfer weight of each source domain.
% was developed to learn an adaptive feature selection for different tasks, by using a shared mixture-of-expert model with task-specific gating networks. 
% It explicitly models task relationships dynamically, allowing for automatically allocating model parameters which alleviates task conflicts in optimization.

Different from \textit{joint training}, \textit{pre-training \& fine-tuning} is a two-stage learning paradigm. In pre-training stage, a model is first trained in a source domain. Then in fine-tuning stage, a target model would load the pre-trained model parameters, and then fine-tune itself for target domain CTR prediction. In each stage, only one objective is used for optimization, and thus the gradient interference issue can be alleviated to some extent. This method has been widely-applied in industrial systems~\cite{chen2021user,liu2022continual,yang2022click,zhang2022keep}. In CTNet~\cite{liu2022continual}, Liu \textit{et al.} focus on the CDCTR problem in a time-evolving scenario. The heterogeneous multi-scenario knowledge transfer problem is studied in~\cite{10.1145/3580305.3599955}. 
Zhang \textit{et al.} proposed KEEP~\cite{zhang2022keep} which is a two-stage framework that consists of a supervised pre-training knowledge extraction module performing on web-scale and long-time source domain data (\textit{i.e.} super domain), and a plug-in network that incorporates the extracted knowledge into the downstream target fine-tuning model.
Different from KEEP, the proposed CDAnet, explicitly learns knowledge transfer across domains without requiring super-domain data and with context-aware user interest transferring.

\begin{figure*}[t]
\centering
\includegraphics[width=15.0cm]{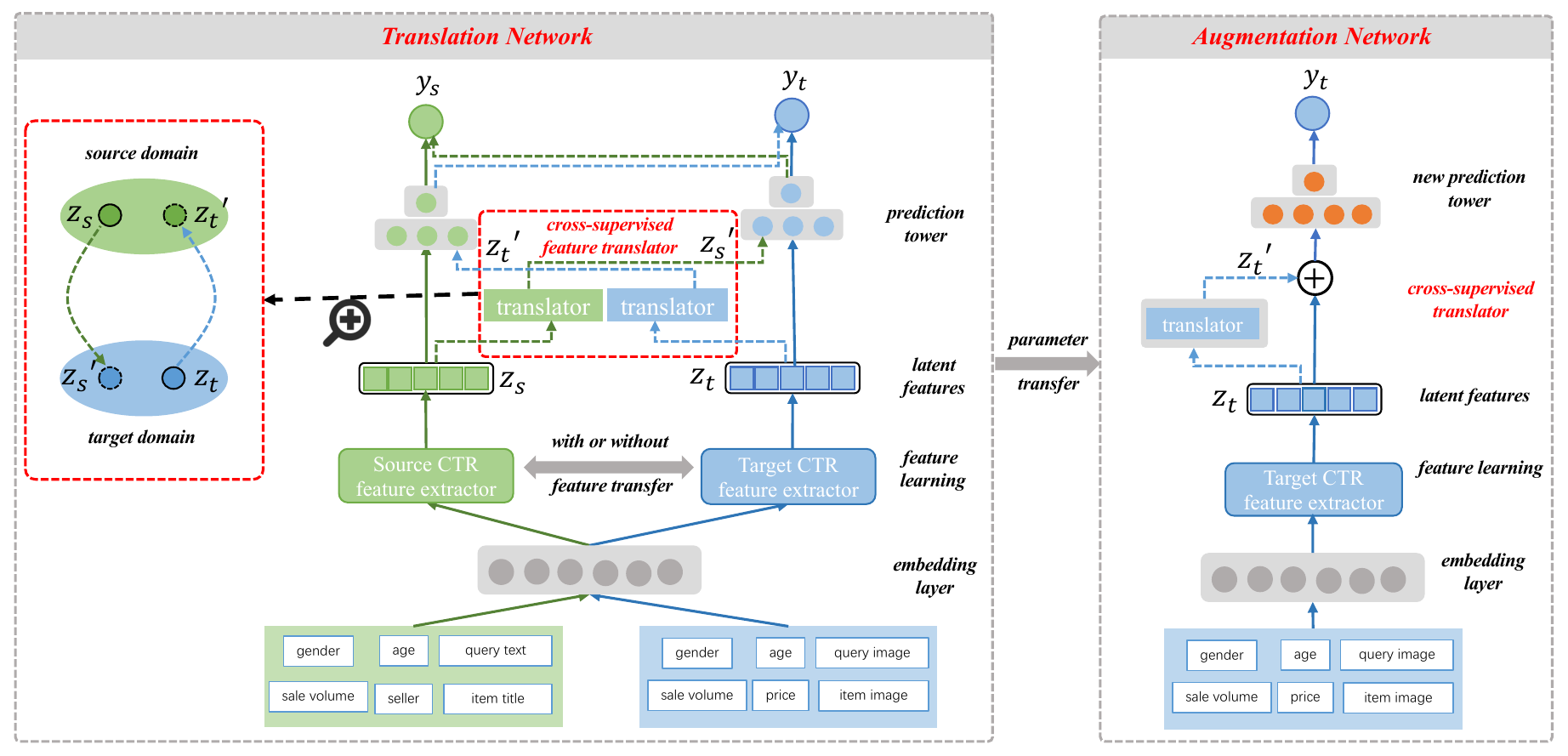}
\vspace{-8pt}
\caption{The architecture of our cross-domain augmentation networks (CDAnet). At the translation stage, the translation network encodes the inputs from two domains as latent features by two feature extractors, and then learns the feature translator. The two extractors could be independent which means no additional feature transfer technique between them. Or we can use some parameter-sharing techniques~\cite{sheng2021one,ouyang2020minet,ma2018modeling,tang2020progressive} to additionally introduce implicit feature transfer ways. The feature translator learns to translate latent feature between two spaces.
% The CDCTR feature learning techniques can be incorporated with existing transfer learning techniques such as shared MLP layers in~\cite{sheng2021one,ouyang2020minet} or advanced mixture-of-experts in~\cite{ma2018modeling,tang2020progressive}.
% Next, the translation network parameters except the tower layer are transferred to the augmentation network. 
At the augmentation stage, translation network parameters except the tower layer are reused. The augmentation network takes the target domain samples and their translated latent features together to boost the target model training. $\oplus$ denotes the concatenation operation. Note that the embedding layer is a simplified plot. When two domains have non-overlapped feature fields, we use different sub-embedding layer.}
\label{figure:model_architecture}
\vspace{-8pt}
\end{figure*}
\section{Method}
\subsection{Preliminary}
In CDCTR prediction, given source domain $\mathcal{S}$, we have its training samples $(\boldsymbol{x_{s}},y_{s})$ where $\boldsymbol{x_{s}} \in \mathbb{R}^{F_{s}\times 1}$ denotes the input features and $y_{s}\in \{0,1\}$ is the click label (\textit{i.e.} 1 means click while 0 means non-click).
Similarly, we have the target domain $\mathcal{T}$ and its training samples $(\boldsymbol{x_{t}},y_{t})$ in which $y_{t}\in \{0,1\}$ is the click label in target domain and $\boldsymbol{x_{t}}\in \mathbb{R}^{F_{t}\times 1}$ is the input features. $F_{s}$ and $F_{t}$ denote the input feature dimension of source and target domain, respectively. The goal is to transfer knowledge from source domain to target domain and improve the target domain CTR prediction.
% We define $x_{s}$ and $x_{t}$ are heterogeneous when they have varying feature fields. For instance, the text query is important and widely used in text2product retrieval. While in image2product retrieval, we do not have the text query but have the image query and the image feature is quite different from the text. When two domains have heterogeneous input features, many recent works cannot well handle this and it is challenging to bridge the heterogeneous gap for efficient knowledge transfer. 
% Moreover, it is nearly impossible to guarantee the same user buy two similar items in two in a practical scenario, 
% In this case,  The goal is to perform knowledge transfer between $\mathcal{S}$ and $\mathcal{T}$, and improve the CTR prediction in target domain.

% The key of modeling in CDCTR with heterogeneous input features is to make sure that the model can embed the heterogeneous inputs and meanwhile have a good ability of transferring knowledge, even without the widely used shared embedding layer technique used in recent works~\cite{ouyang2020minet,liu2022continual,ma2018modeling}. Accordingly, we propose our 
In the proposed CDAnet, there are two sequentially learned networks: translation network and augmentation network.
First, the translation network encodes the inputs as latent features and learns how the latent features are translated by a designed cross-supervised feature translator. Then, the pre-trained parameters of translation network are transferred to the augmentation network. Next, the augmentation network will combine the translated latent features and the original latent features of target domain samples together to augment the target domain model training. The model architecture is shown in Figure~\ref{figure:model_architecture}. Details about each module are demonstrated in the following parts.

\subsection{Translation Network}
The translation network aims to learn latent feature translation between source domain and target domain. It has four parts including embedding layer, feature extractor, cross-supervised feature translator and prediction tower. 

\textbf{Embedding Layer}: The embedding layer aims to encode the various input features from different domains as embeddings.
% Since the input features are heterogeneous, it is necessary to decouple the embedding layer of two domains so that the embedding layer can adapt to the input patterns in its own domain. 
In the embedding layer, overlapped feature fields have shared sub-embedding parts while non-overlapped feature fields use non-shared sub-embedding parts. We briefly denote the the whole embedding layer as $E$ for simplicity.
Given the inputs $\boldsymbol{x_{s}}$ of source domain and $\boldsymbol{x_{t}}$ of the target domain, we have:
\begin{align}
\boldsymbol{e_{s}}=E(\boldsymbol{x_{s}}),~\boldsymbol{e_{t}}=E(\boldsymbol{x_{t}})
\end{align}
where $\boldsymbol{e_{s}}\in \mathbb{R}^{d\times 1}$ and $\boldsymbol{e_{t}}\in \mathbb{R}^{d\times 1}$ are the embedding features of $\boldsymbol{x_{s}}$ and $\boldsymbol{x_{t}}$, respectively. $d$ is the embedding dimension.

\textbf{Feature Extractor}: 
After embedding layer, CTR prediction usually has feature extraction module to capture high-order feature interactions for user behavior modeling. In CDCTR, we similarly have this feature extraction module and it can be fused with some transfer learning techniques such as vanilla shared MLP in~\cite{sheng2021one,zhang2022keep} and mixture-of-expert in~\cite{ma2018modeling,tang2020progressive}
that implicitly transfer knowledge across domains.
The vanilla shared MLP has the advantage of simplicity and low computation complexity but may have the risk of gradient interference issue when optimizing different objectives. The mixture-of-expert technique can alleviate this gradient interference issue but has higher computation complexity. Our CDAnet is flexible on the choice of feature extractor such as shared MLP or mixture-of-expert, which means it could be combined with existing implicit knowledge transfer techniques.
To make an easy demonstration, we denote the feature extractor of source domain and target domain as $F_{s}$ and $F_{t}$, respectively. Then, we have:
\begin{align}
    \boldsymbol{z_{s}}=F_{s}(\boldsymbol{e_{s}}),~\boldsymbol{z_{t}}=F_{t}(\boldsymbol{e_{t}})
\end{align}
where $\boldsymbol{z_{s}}\in \mathbb{R}^d$ and $\boldsymbol{z_{t}}\in \mathbb{R}^d$ are extracted latent features of source domain and target domain, respectively. Note $F_{s}$ and $F_{t}$ could be independent, \textit{i.e.} no transfer technique between $F_{s}$ and $F_{t}$, or they could be related
by some transfer techniques such as MMOE~\cite{ma2018modeling}.

\textbf{Cross-supervised Feature Translator}: After obtaining the latent features of each domain, we propose a cross-supervised translator to learn the latent feature translation, which is inspired by image2image translation~\cite{pang2021image,isola2017image,liu2017unsupervised,yi2017dualgan,zhu2017unpaired,kim2017learning}. 
In image2image translation, images from target domain are translated as the images in the source domain, supervised by the true images in source domain. 
Whereas in CDCTR, the sample (\textit{i.e.} combined user feature and item feature) in the target domain should not share the user behavior label with a sample in the source domain, because these two samples do not have paired relationship.

\underline{\emph{Remark.}} Instead, a widely held belief in CDCTR is that if a user favors an item in the target domain, the favor behavior is preserved if the user and item are mapped into the source domain as corresponding features. For example, if a user likes science movies, he or she will also tend to love science novels. This indicates that a target domain sample should preserve its target domain label when this sample is translated as corresponding features in source domain.

% That is to say, we cannot have paired supervision of two samples   

% the combination of a user feature and item feature from target domain may not share the label in source domain, because . 
% For example, when a user clicked an item in target domain, we believe if the user and item are mapped into the source domain as corresponding features, the click behavior preserves. For example, if a user likes science movies, he or she will also tend to love science novels. This is a basic belief in CDCTR and has been widely employed in recent works~\cite{yuan2019darec,chen2020towards}. 

In other words, given the latent feature $\boldsymbol{z_{t}}$ of target domain, we aim to translate it into source domain as $\boldsymbol{z_{t}^{'}}$ while preserving the content of $\boldsymbol{z_{t}}$. Then, $\boldsymbol{z_{t}^{'}}$ can be taken into the prediction tower of source domain while supervised by the target domain label $y_{t}$.
Let $\boldsymbol{W_{t}^{tran}}\in \mathbb{R}^{d\times d}$ be the translator of target domain and $BCE$ be the binary cross entropy loss, then the translator of target domain is optimized by:
\begin{align}
\label{eq:L_t_cross}
    \min~\mathcal{L}_{t}^{cross}=BCE(\sigma(R_{s}(\boldsymbol{z_{t}^{'}})),y_{t}),~~\boldsymbol{z_{t}^{'}} = \boldsymbol{W_{t}^{tran}}\boldsymbol{z_{t}}
\end{align}
where $\sigma$ is the \text{sigmoid} function and $R_{s}$ is the prediction tower of source domain that maps latent feature into prediction logit. 
To stabilize the training, we have the symmetric formulation for source domain like Eq.~\ref{eq:L_t_cross} as:
\begin{align}
    \min~\mathcal{L}_{s}^{cross}=BCE(\sigma(R_{t}(\boldsymbol{z_{s}^{'}})),y_{s}),~~\boldsymbol{z_{s}^{'}} = \boldsymbol{W_{s}^{tran}}\boldsymbol{z_{s}}
\end{align}
where $R_{t}$ is the prediction tower of target domain and $\boldsymbol{W_{s}^{tran}}\in \mathbb{R}^{d\times d}$ is the translator of source domain. The network architecture of $R_{s}$ and $R_{t}$ is the commonly used MLP layers.
% could be different architectures according to the domain's own characteristics. In our experiment, we simply use  MLP 

In order to better conduct the latent feature translation, apart from the above cross supervision, we add an orthogonal mapping constraint on the translators $\boldsymbol{W_{s}^{tran}}$ and $\boldsymbol{W_{t}^{tran}}$ as:
{\small
\begin{align}
\label{eq:orth_loss}
    \min~\mathcal{L}^{orth}=&||\mathbb{T}(\boldsymbol{W_{s}^{tran}})(\boldsymbol{W_{s}^{tran}}\boldsymbol{z_{s})-z_{s}}||_{F}^{2}+ \nonumber \\
    &||\mathbb{T}(\boldsymbol{W_{t}^{tran}})(\boldsymbol{W_{t}^{tran}}\boldsymbol{z_{t}})-\boldsymbol{z_{t}}||_{F}^{2}
\end{align}
}%
where $\mathbb{T}$ is the transpose operation. The orthogonal transformation in mathematics has a characteristic that can preserve lengths and angles between vectors. Therefore, the orthogonal mapping constraint in Eq.~\ref{eq:orth_loss} can help the latent features $\boldsymbol{z_{t}}$ of different samples preserve the similarity and avoid a case where multiple $\boldsymbol{z_{t}}$ collapse to a single point after translation.

\subsubsection{Objective Function in Translation}
Apart from the above objective of learning translator, we still need the vanilla objective for optimizing the CTR task in each domain. Namely, the vanilla CTR objectives of source and target domain are formulated as:
{\small
\begin{align}
    \min \mathcal{L}_{s}^{vani}=BCE(\sigma(R_{s}(\boldsymbol{z_{s}})),y_{s}),
    \mathcal{L}_{t}^{vani}=BCE(\sigma(R_{t}(\boldsymbol{z_{t}})),y_{t})
\end{align}
}%
\underline{\emph{Remark}.} These two objectives help the model learn the network parameters (\textit{e.g.} the tower network) targeted for CTR prediction. Meanwhile, when the tower networks are optimized for each domain's CTR prediction, the translators can adapt to the tower networks and learn feature translation in a meaningful and right direction. To sum up, the objective of translation network is:
\begin{align}
\label{eq:translation_objective}
    \min \mathcal{L}_{trans}=\underbrace{\mathcal{L}_{s}^{vani}+\mathcal{L}_{t}^{vani}}_{\mathcal{L}^{vani}}+\alpha (\underbrace{\mathcal{L}_{s}^{cross}+\mathcal{L}_{t}^{cross}}_{\mathcal{L}^{cross}})+\beta \mathcal{L}^{orth}
\end{align}
where $\alpha$ and $\beta$ are hyper-parameters on loss weights. 

\subsection{Augmentation Network}
The augmentation network exploits the previously learned feature translator to augment the target model's fine-tuning for boosted CTR performance.
In particular, we would first transfer the network parameters including the embedding layer, the feature extractor module and the translator to the augmentation network. As shown in Figure~\ref{figure:model_architecture}, in order to enable learning flexibility for augmentation network and adapt to the feature shift between two stages,  the prediction tower is newly initialized rather than coming from the translation network. Then, given a target domain training sample, we can combine its original latent feature $\boldsymbol{z_{t}}$ and the additional translated latent feature $\boldsymbol{z_{t}^{'}}$ to conduct cross-domain augmentation for the target model's fine-tuning.

\textbf{Cross-domain Augmentation}: In cross-domain augmentation, the augmented latent feature of target domain is formulated as:
\begin{align}
    \boldsymbol{z_{t}^{aug}}=\boldsymbol{z_{t}}\oplus \boldsymbol{z_{t}^{'}}
\end{align}
where $\oplus$ denotes the concatenation operation. 
% Other combination ways of $\boldsymbol{z_{t}}$ and $\boldsymbol{z_{t}^{'}}$ are investigated in experiments. 
Also, when caring about the augmentation in source domain, the augmented feature can be obtained in a similar formula.

\textbf{Objective Function in Augmentation}
With the augmented latent feature $\boldsymbol{z_{t}^{aug}}$, we would feed it into the new prediction tower $R_{t}^{aug}$ and use the vanilla CTR objective for fine-tuning. The objective is defined as:
\begin{align}
    \mathcal{L}_{aug}=BCE(\sigma(R_{t}^{aug}(\boldsymbol{z_{t}^{aug}})),y_{t})
\end{align}
In this augmentation stage, the model has only one CTR objective and can avoid the optimization conflict problem in multi-objective models. When focusing on the performance of source domain, its augmentation network can be optimized in a similar way. The constraint in Eq.~\ref{eq:orth_loss} also works here with the same $\beta$ as in Eq.~\ref{eq:translation_objective}.\\

\noindent\underline{\emph{Discussion.}}
Considering the technique in knowledge transfer, the \textit{joint training} works~\cite{hu2018conet,ma2018modeling,tang2020progressive,sheng2021one} and fine-tuning
methods~\cite{ouyang2020minet,liu2022continual,zhang2022keep} mainly employ different parameter-sharing techniques to regularize the target model training~\cite{hu2018conet,li2019ddtcdr,yuan2019darec,ma2018modeling,zhang2022keep} and implicitly learn the knowledge from source domain. 
To achieve more efficient knowledge transfer, KEEP~\cite{zhang2022keep} introduces a way to explicitly extract the knowledge from source domain, but it is limited in requiring super-domain data and learning static user interests. By contrast, the proposed CDAnet explicitly learns knowledge transfer by a novel translation then augmentation idea. It has \textbf{no super-domain data requirement} and enables \textbf{context-aware} user interest transfer, which shows its \emph{flexibility} and \emph{efficiency}. Moreover, CDAnet is compatible with the parameter-sharing technique to achieve more flexible usage and better performance in different scenarios.
% We notice that most of the above works employ different parameter-sharing technique to regularize the target model training~\cite{hu2018conet,li2019ddtcdr,yuan2019darec,ma2018modeling,zhang2022keep}. This regularization can partly prevent the target model training from over-fitting suspicious features, and thus implicitly learn the knowledge from source domain. Although KEEP~\cite{zhang2022keep} introduces a way to explicitly use the knowledge from source domain, it is limited in requiring super-domain data and learning static user interests.
% By contrast, the proposed CDAnet, explicitly learns knowledge transfer across domains and is more flexible without requiring super-domain data.

% In addition, CDAnet supports heterogeneous input features while most existing CDCTR models require homogeneous inputs.
% CDAnet inherits the advantage of separate training as \textit{pre-training \& fine-tuning} 
% and includes a novel translation then augmentation idea for knowledge transfer rather than the shared embedding layer. Further, different from existing \textit{pre-training \& fine-tuning} method that requires homogeneous input features, our method enables heterogeneous inputs and is more flexible in real-world applications.

\begin{table}[]
\centering
\caption{The statistics of datasets.}
\vspace{-8pt}
\label{table:dataset}
\renewcommand{\arraystretch}{0.8}
 \setlength{\tabcolsep}{0.7mm}{ 
  \scalebox{0.85}{
\begin{tabular}{ccccc}
\hline
dataset               & \multicolumn{2}{c}{Amazon(movie and book)} & \multicolumn{2}{c}{Taobao(ad and rec)} \\ \hline
domain                & movie                & book                 & ad                 & rec                \\ \hline
\#users               & 29,680               & 52,690               & 141,917            & 186,731            \\
\#items               & 16,494               & 47,302               & 165,689            & 379,817            \\
\#input feature dim & 8,044                & 24,466               & 44,897             & 5,004              \\
\#train samples        & 1,685,836            & 7,133,107            & 3,576,414          & 12,168,878         \\
\#val samples              & 210,730              & 891,638              & 447,052            & 1,521,110            \\
\#test samples              & 210,730              & 891,639              & 447,052            & 1,511,110            \\
\#positive samples              & 351,216              & 1,486,064              & 234,736            & 855,362 \\
% \#sparsity              & 99.92\%              & 99.94\%              & 99.99\%            & 99.99\% \\ 
\hline
\end{tabular}
}}
\vspace{-12pt}
\end{table}
\section{Experiments and Analysis}
\subsection{Experiment Setup}
\subsubsection{Datasets}
\label{sec:datasets}
We first conduct our experiments on two public benchmarks:
Amazon\footnote{https://jmcauley.ucsd.edu/data/amazon/} and Taobao\footnote{https://tianchi.aliyun.com/dataset/56}\footnote{https://tianchi.aliyun.com/dataset/649}.
For Amazon, we choose two largest domains--movie and book to conduct experiments. In movie domain, we have user ID, movie ID, movie genre, movie director and movie name information. While in book domain, we have user ID, book ID, book category, book writer and book name information.
The original user behaviors are 0-5 ratings and we process ratings larger than 3 as positive feedback and others as negative feedback for CTR prediction. Both movie and book name are processed as vectors by a word2vec Glove-6B model\footnote{https://nlp.stanford.edu/projects/glove/}. 
% The movie genre, movie director, book category and book writer are mapped as one-hot features, respectively for each domain.
Taobao dataset contains user-item interactions of advertisement (ad) and recommendation (rec) domain. We consider buy behavior type as positive feedback and no-buy as negative feedback. 
In ad domain, we get user ID, age, gender, occupation, some other user profile information, ad ID, ad category and ad brand information. 
In rec domain, we get user ID, item ID and item category information.
For both datasets, user and item ID are both embedded as 64-dim features. Other discrete features are processed as one-hot or multi-hot features. 
% For both datasets, we also apply a $k$-core filtering to guarantee each user or item has at least $k$ interactions. $k$ is 5 and 10 for movie and book domain of Amazon, respectively. For Taobao, $k$ also equals 5 and 10 for ad and rec domain, respectively. 
The dataset statistics are summarized in Table~\ref{table:dataset}.

Furthermore, we also evaluate our model on Alibaba production data and online experiments on Taobao mobile app. In particular, we collect six-month
% (from 2022/06/20 to 2022/12/20)
user behaviors in text2product retrieval as source domain data and one-year 
% (from 2021/12/20 to 2022/12/20) 
user behaviors in image2product retrieval as target domain data.
% and meanwhile remove data of special holidays. Data from 2022/12/21 to 2022/12/25 is taken as the offline test set. 
The number of train data in text2product is nearly four times larger than that of image2product domain. Both domains contain hundreds of billions of samples and hundreds of input feature fields that are used in our online production system. 
% The input features in these two domains are quite different due to the characteristic of these scenarios.

\subsubsection{Baselines}
We adopt different strong algorithms for comparison, including the single domain, \textit{joint training} and \textit{pre-training \& fine-tuning} methods. The single domain models are: 1) \textbf{MLP}. A deep multi-layer perception (MLP) model is a common and efficient single-domain ranking model in online search and recommendation systems. 2) \textbf{ShareBottom}. It is a multi-task model that shares parameters of the bottom layers. In our implementation, since two domains may have non-overlapped feature fields, we thus only share the embedding parameters of overlapped feature fields and two middle-layer feature mapping parameters. The \textit{joint training} models are: 3) \textbf{STAR}~\cite{sheng2021one}. STAR is a star topology model that trains a single model to serve all domains by leveraging the data from all domains simultaneously.
4) \textbf{DDTCDR}~\cite{li2019ddtcdr}. It is a deep dual transfer learning model that transfers knowledge between related domains in an iterative manner.
5) \textbf{MMOE}~\cite{ma2018modeling}. MMOE implicitly models task relationships for multi-task learning by a shared mixture-of-experts module and task-specific gates. 
6) \textbf{PLE}~\cite{tang2020progressive}. PLE is a multi-task learning model that separates shared components and task-specific components and adopts a progressive routing mechanism to extract and transfer knowledge. The \textit{pre-training \& fine-tuning} method: 7) \textbf{KEEP}~\cite{zhang2022keep}. KEEP is an industrial knowledge extraction and plugging framework for online recommendation\footnote{We only use it on the production dataset because the public benchmarks do not have a super domain that covers most users or items of the target domain}.

\subsubsection{Parameter Settings}
In our experiments on public benchmarks, we split the data into train, validation, and test sets with the common 8:1:1 setting according to chronological order. The experiments are conducted multiple times and the mean value is taken as the model performance. We set the latent feature size as 128 for all models. We use the validation performance as early stop condition and the max number of training epoch is 200.
% which can ensure the model's convergence. 
To make a fair comparison, we conduct a grid search of hyper-parameters and the number of layers for all models. On Amazon dataset, $\alpha$ is 0.01, $\beta$ equals 0.1, the number of experts is 2 and each expert has 2 layers when using mixture-of-experts, the prediction tower has 2 layers including the logit mapping layer.
On Taobao dataset, $\alpha$ is 0.03, $\beta$ equals 0.1, the number of experts is 2 and each expert has 2 layers when using mixture-of-experts, and the prediction tower has 3 layers including the logit mapping layer.

% The two domains have their own embedding layer and a shared 2-layer and 2-expert MMOE module. The hyper-parameter $\alpha$ and $\beta$ are both 0.01. Note that the hyper-parameters of industrial dataset are set based on the experience of public benchmarks

\begin{table}[]
\centering
\caption{AUC comparison results of different models on Amazon and Taobao. 
% The results of ``source$\rightarrow$target'' domain are listed. ``CDAnet'' indicates the feature extractors across domains are independent and there is no parameter-sharing between the two extractors. 
``CDAnet+IndepMLP'' indicates the feature extractor of both domains are independent MLP. ``CDAnet+SharedMLP/PLE/MMOE'' means we take shared MLP/PLE/MMOE as the feature extraction module.}
\vspace{-6pt}
\label{table:overall_public}
\renewcommand{\arraystretch}{0.85}
 \setlength{\tabcolsep}{0.5mm}{ 
  \scalebox{0.9}{
\begin{tabular}{ccccc}
\hline
Dataset     & \multicolumn{2}{c}{Amazon} & \multicolumn{2}{c}{Taobao}                                                      \\ \hline
Source$\rightarrow$Target      & Book$\rightarrow$Movie            & Movie$\rightarrow$Book    & Rec$\rightarrow$Ad                                     & Ad$\rightarrow$Rec                                    \\ \hline
MLP         & 0.6595           & 0.7604  & 0.6161                                 & 0.6865                                 \\
ShareBottom & 0.6604           & 0.7603  & 0.6160                                  & 0.5020                                  \\
STAR        & 0.6503           & 0.7626  & 0.6149                                 & 0.6795                                 \\
DDTCDR      & 0.6636           & 0.7807  & 0.6162                                 & 0.6756                                 \\
MMOE        & 0.7025           & \textbf{0.7899}  & 0.6177                                 & 0.6930                                 \\
PLE         & 0.6993           & 0.7846  & 0.6164                                 & 0.6933                                 \\ \hline
CDAnet+IndepMLP       & 0.7207  &  0.7735       & \textbf{0.6205} & \textbf{0.7047} \\
CDAnet+SharedMLP       & 0.7203  &  0.7847       & 0.6195 & 0.7034 \\
CDAnet+PLE       & 0.7213  &  0.7838       & 0.6201 & 0.7023 \\
CDAnet+MMOE       & \textbf{0.7225}  &  0.7811       & 0.6200 & 0.7034 \\ \hline
\end{tabular}
}}
\vspace{-8pt}
\end{table}

% \begin{table}[]
% \centering
% \caption{The AUC comparison results of Pailitao domain on our industrial dataset. MainSe is the source domain of text2product retrieval and Pailitao is our target domain of image2product retrieval.}
% \label{table:overall_industrial}
% \begin{tabular}{cc}
% \hline
% Domain & MainsSe~\rightarrow Pailitao \\ \hline
% Base   & 0.7734                                         \\
% CDAnet  & -                                         \\ \hline
% \end{tabular}
% \end{table}

\begin{table}[]
\centering
\caption{Overall comparison results of different models on our image2product domain. Here we consider text2product as the source domain and image2product domain as the target domain. AUC indicates the offline evaluation metric. CTR, deal number and GMV are online A/B test metrics. Due to the company's regulations, some online metric values of Base model are blinded and denoted as $\star$. We provide the improvement gap of CDAnet.}
\vspace{-6pt}
% \vspace{-8pt}
% Since our production model of CTR and CVR task are two different models, we apply CDAnet on both tasks and report the performance.
\label{table:overall_industrial}
\renewcommand{\arraystretch}{1.0}
\setlength{\tabcolsep}{0.8mm}{ 
\scalebox{0.9}{
\begin{tabular}{ccccc}
\hline
Model  & AUC                                       & CTR        & deal number & GMV     \\ \hline
Base   & 0.7845                                    & 9.50\%          & $\star$           & $\star$       \\
KEEP   & 0.7858(+0.13\%)                           & -          & -           & -       \\ \hline
CDAnet & \textbf{0.7866(+0.21\%)} & \textbf{9.61\%(+0.11point)} & \textbf{+0.64\%}     & \textbf{+1.26\%} \\ \hline
\end{tabular}
}}
\vspace{-8pt}
\end{table}

\begin{table}[]
\centering
\caption{Online CTR improvement gap of CDAnet on different views. ``User Group'' indicates users are split into different groups according to their consumption power. ``Query Category'' indicates the query image category in our business. The improvement gap is measure by ``point'' as it in Table~\ref{table:overall_industrial}.}
\vspace{-6pt}
% \vspace{-8pt}
% Since our production model of CTR and CVR task are two different models, we apply CDAnet on both tasks and report the performance.
\label{table:finegrained_industrial}
\renewcommand{\arraystretch}{1.0}
\setlength{\tabcolsep}{0.5mm}{ 
\scalebox{0.9}{
\begin{tabular}{c|cccccccccc}
\hline
% View    & \multicolumn{10}{c}{Price Level}                                                                                                           \\ \hline
% Type    & \multicolumn{2}{c}{L1}    & \multicolumn{2}{c}{L2}    & \multicolumn{2}{c}{L3}    & \multicolumn{2}{c}{L4}    & \multicolumn{2}{c}{L5}    \\ \hline
% Improv. & \multicolumn{2}{c}{+0.23} & \multicolumn{2}{c}{+0.14} & \multicolumn{2}{c}{+0.04} & \multicolumn{2}{c}{+0.02} & \multicolumn{2}{c}{+0.04} \\ \hline \hline
View    & \multicolumn{10}{c}{User Group}                                                                                                           \\ \hline
Type    & \multicolumn{2}{c}{G1}    & \multicolumn{2}{c}{G2}    & \multicolumn{2}{c}{G3}    & \multicolumn{2}{c}{G4}    & \multicolumn{2}{c}{G5}    \\ \hline
Improv. & \multicolumn{2}{c}{+0.09} & \multicolumn{2}{c}{+0.11} & \multicolumn{2}{c}{+0.13} & \multicolumn{2}{c}{+0.13} & \multicolumn{2}{c}{+0.12} \\ \hline \hline
View    & \multicolumn{10}{c}{Query Category}                                                                                                             \\ \hline
Type    & C1           & C2         & C3           & C4         & C5           & C6         & C7           & C8         & C9           & C10        \\ \hline
Improv. & +0.14        & +0.14      & +0.08        & +0.12      & +0.15        & +0.13      & +0.16        & +0.15      & +0.12        & +0.14      \\ \hline
\end{tabular}
}}
\vspace{-8pt}
\end{table}

\subsection{Overall Comparison}
% \subsubsection{Offline Comparison}
% the feature extractor module is taken as both no-parameter-sharing (\textit{i.e.} IndependentMLP) and parameter-sharing techniques (\textit{i.e.} SharedMLP, MMOE and PLE) to study model performance under different situations.
In this part, we make model comparisons with various baselines on different public benchmarks. Meanwhile we study the model performance when the feature extractor is combined with no-parameter-sharing (\textit{i.e.} CDAnet+IndepMLP) and existing implicit knowledge transfer techniques (\textit{i.e.} CDAnet+SharedMLP/PLE/MMOE). 
% The results are summarized in Table~\ref{table:overall_public}.
% In Table~\ref{table:overall_public}, ``CDAnet'' means feature extractors of two domains are independent MLP.
% ``CDAnet+SharedMLP/MMOE/PLE'' indicates different implicit knowledge transfer techniques are taken in the design of feature extractors. 

From Table~\ref{table:overall_public}, we observe that CDAnet with different feature extractors generally achieve better performance than other models. First, the \textit{joint training} models (\textit{i.e.} ShareBottom, DDTCDR, MMOE, PLE) mainly rely on shared embedding layers for implicit knowledge transfer. Whereas in most real-world cases, there are not many overlapped feature fields across domains and the used benchmarks are in this case. Then, most embedding parameters of these models cannot be shared and the knowledge transfer is largely affected. By contrast, CDAnet relies on the explicitly translated knowledge to augment the target model for transferring, and thus can keep good performance. Second, STAR mainly relies on a centered and domain-shared network to transfer knowledge, which is much less efficient than CDAnet's translation and augmentation idea for explicit knowledge transfer. Third, we see that ``CDAnet+IndepMLP'' achieves better performance than ``CDAnet+SharedMLP/MMOE/PLE'' on Tabao dataset while not on Amazon dataset. Combining the implicit knowledge transfer techniques with CDAnet on one hand could regularize the model training, but on the other hand may limit the ability of fitting more useful features. The key is which part takes the main position in model learning, and this is usually determined by model architectures and data scenarios. With CDAnet, we can flexibly choose feature extractors to adapt to the data scenarios and achieve well performance.
\begin{table}[]
\centering
\caption{CDAnet's relative improvement gap regarding the exposed product quality (EPQ) score and the deal number at different item price levels. 
% The item value level is determined based on the item price. Generally speaking, the higher price is, the high value is.
(a) shows CDAnet's relative improvement gap on the EPQ score and (b) indicates the relative improvement gap on deal number.}
\vspace{-6pt}
% \vspace{-8pt}
% Since our production model of CTR and CVR task are two different models, we apply CDAnet on both tasks and report the performance.
\label{table:improved_gap}
\renewcommand{\arraystretch}{1.0}
\setlength{\tabcolsep}{1.0mm}{ 
\scalebox{1.0}{
\begin{tabular}{cccccc}
\hline
Price Level   & L1     & L2     & L3     & L4     & L5     \\ \hline
Rel. Improv. EPQ score(\%) & +2.10 & +3.62 & +3.60 & +3.43 & +2.68 \\
Rel. Improv. of deal(\%)   & +0.08  & +0.12 & +0.11 & +0.15 & +0.17 \\ \hline
\end{tabular}
}}
\vspace{-8pt}
\end{table}

% \begin{figure}[t]
% \centering
% \begin{minipage}[t]{0.23\textwidth}
% \centering
% \includegraphics[width=\textwidth]{images/seller_star_improvement_gap.png}
% \vspace{-20pt}
% \caption*{(a) \footnotesize{Exposed product quality score}}
% \end{minipage}
% \begin{minipage}[t]{0.23\textwidth}
% \centering
% \includegraphics[width=\textwidth]{images/pay_improvement_gap.png}
% \vspace{-20pt}
% \caption*{(b) \footnotesize{Deal number}}
% \end{minipage}
% \vspace{-6pt}
% \caption{CDAnet's improvement gap regarding the exposed product quality (EPQ) score and the deal number at different item price levels. 
% % The item value level is determined based on the item price. Generally speaking, the higher price is, the high value is.
% (a) shows CDAnet's improvement gap on the EPQ score and (b) indicates the improvement gap on deal number.}
% \label{figure:improved_gap}
% % \vspace{-10pt}
% \end{figure}

\begin{figure*}[t]
\centering
\begin{minipage}[t]{0.46\textwidth}
\centering
\includegraphics[width=\textwidth]{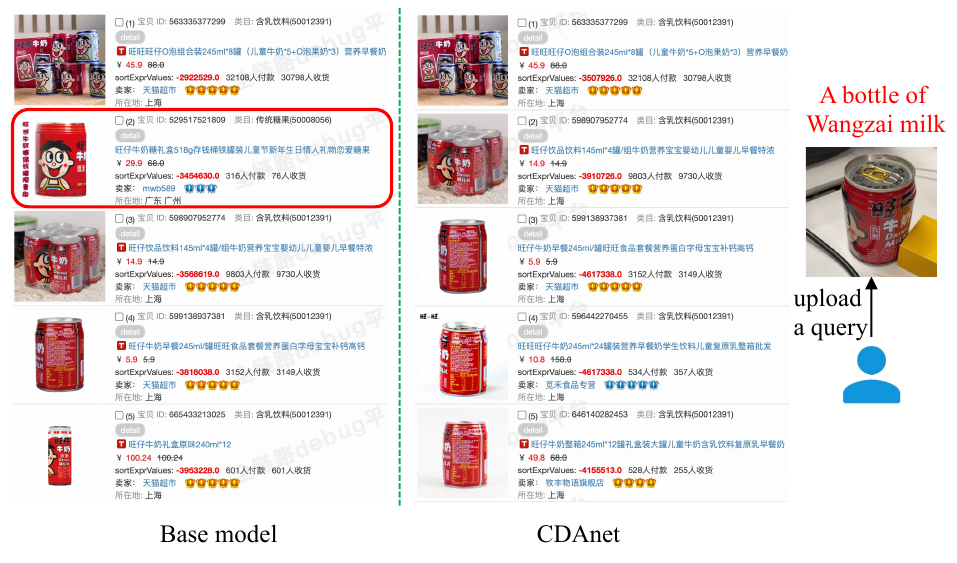}
\vspace{-23pt}
\caption*{(a) \footnotesize{example of Wangzai milk}}
\end{minipage}
\hspace{5pt}
\begin{minipage}[t]{0.46\textwidth}
\centering
\includegraphics[width=\textwidth]{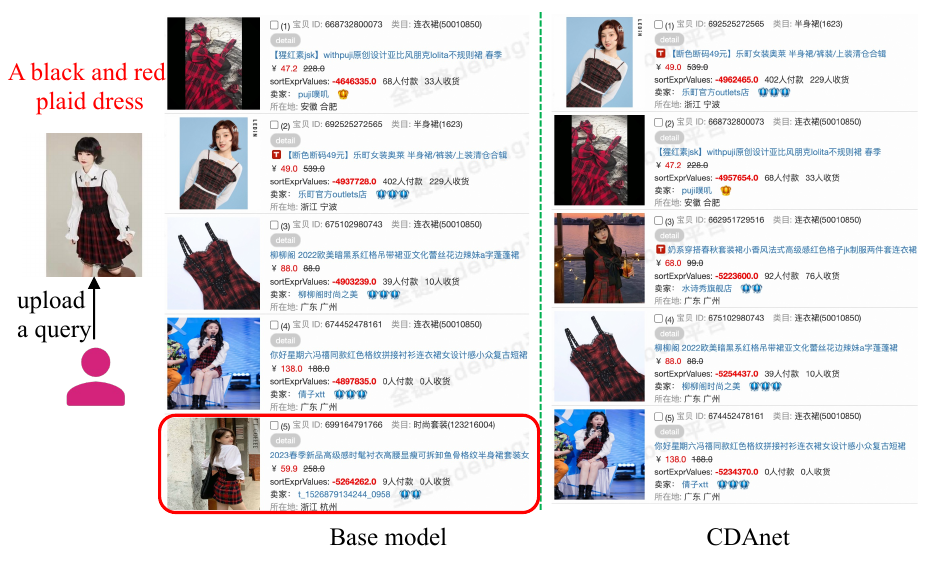}
\vspace{-23pt}
\caption*{(b) \footnotesize{example of a dress}}
\end{minipage}
\vspace{-8pt}
\caption{Ranking examples of online Base model and our CDAnet. Given an image query, we list the top5 ranking items of different models. (a) is an example of Wangzai milk and (b) is an example of a dress. Bad cases of Base model are circled with red box. In (a), the user expected to find a bottle of Wangzai milk, but Base model returned a bad case which is a bottle of Wangzai candy. In (b), the user uploaded a query of black and red plaid dress, but Base model returned a bad case which is a skirt.}
\label{figure:ranking_exmples}
\vspace{-8pt}
\end{figure*}

\subsection{Production Deployment}
We further evaluate CDAnet through production deployment in image2product retrieval of Alibaba Taobao system. In the deployment, the training dataset is previously illustrated in Section~\ref{sec:datasets}. Since the text2product retrieval and our image2product retrieval have many non-overlapped feature fields, we only partly share embedding layer parameters of two domains.
Then, after embedding layer, we empirically use DCN-v2~\cite{wang2021dcn} to extract latent features of text2product domain\footnote{We have no access to the online production model of text2product domain.}, while feature extractor of our image2product domain is our online serving model. The prediction tower of both domains is two-layer MLP including the logit mapping layer. Based on the experience on public benchmarks, we set $\alpha$ and $\beta$ as 0.01.
% We do not tune the hyper-parameters on this production dataset since it would cost too much computation resource. 
% Based on the experience on public benchmarks, we set $\alpha$ and $\beta$ as 0.01.

\subsubsection{Offline Performance on Production Dataset}
In this part, we implement KEEP~\cite{zhang2022keep}, which is one state-of-the-art cross-domain knowledge transfer algorithm for large-scale industrial system, as the comparison model. It takes the text2product domain as super-domain and image2product domain as target domain.
The comparison results are summarized in Table~\ref{table:overall_industrial}.
KEEP explicitly extracts user interests from text2product domain and provides this information for our image2product domain. However, it only provides user interests while ignoring other various features, and the extracted user interests are static no matter what the context features of the target domain inputs are. Instead, CDAnet does not have the above drawbacks and performs better. It brings an absolute 0.21\% AUC increase, which is significant for Taobao's image2product retrieval.

\subsubsection{Online A/B Test}
We conducted online experiments in an A/B test framework of image2product retrieval at Taobao app during the period from 2022/12/26 to 2023/01/16 and from 2023/02/06 to 2023/03/31. In this experiment, the baseline model is our online serving model trained only on image2product domain data. The online evaluation metrics are real CTR, deal number and GMV. The result in Table~\ref{table:overall_industrial} shows that CDAnet leads to an absolute \textbf{0.11 point} CTR increase, a relative \textbf{0.64\% }deal growth and a relative \textbf{1.26\%} GMV increase. In Table~\ref{table:finegrained_industrial}, we also listed CDAnet's CTR improvements gap over online base model regarding different user groups and query categories to more comprehensively evaluate model performance. From this table, we see that CDAnet has obvious improvement under different evaluation views, demonstrating its effectiveness in our production system. It has been fully deployed online since April 2023, serving hundreds and millions of consumers.
% This result shows that CDAnet has the better ranking ability and even can improve the data self-circulation problem in our image2product retrieval system. 

It is also interesting to find that the GMV increase is much larger than the deal growth. This indicates that the average deal price becomes larger and we want to investigate the reason behind it. Specifically, the items are divided into 5 different price levels from low to high according to their price. Then, we investigate how CDAnet performs on these items with different price levels. In Table~\ref{table:improved_gap}, compared to our online base model, we show the improvement gap of CDAnet in terms of the exposed product quality(EPQ) score\footnote{This is a designed score in our business to reflect the quality of the exposed products regarding seller ratings and product sales volume. The larger score is, the higher quality is.} and the deal number at different price levels. From Table~\ref{table:improved_gap}, we see that CDAnet tends to improve the EPQ score much more at high price levels and the deal increase is also larger at high price levels. This result is quite meaningful. In our image2product retrieval, we face a severe data imbalance problem where the high price products (e.g. Beats earphone, iPhone) are less likely searched than the low price ones (e.g. a cup, a shirt), so the user behaviors are too sparse to learn a reliable model on the high price ones. With CDAnet, the rich and huge data knowledge accumulated for many years in text2product retrieval could be transferred to our image2product retrieval. 
As a result, at high price levels of our image2product retrieval, the EPQ score is largely improved, and we could provide better ranking items to encourage deals.

\subsubsection{Online Ranking Example}
% In addition, we verify that CDAnet improves the ranking ability of product-matching in image2product domain by transferring text-matching knowledge from text2product retrieval. 
In addition, we compare some online ranking examples between the baseline model and CDAnet to empirically show the improved ranking ability. The result is shown in Figure~\ref{figure:ranking_exmples}.
% In this figure, we give two examples to illustrate the better ranking ability of CDAnet. 
In (a), the user uploaded a picture of Wangzai milk, we see the online Base model ranks a bottle of Wangzai candy (circled with red box) that is quite visually similar but not the expected product in position 2. Instead, CDAnet can exclude this case and the top5 items are all about Wangzai milk. A possible reason may be that the Wangzai milk and candy in text domain are quite different, while they are not easily to be distinguished in image domain. 
CDAnet can transfer the knowledge of text2product domain into the image2product domain and improve the ranking results.
% The CTR model in text2product retrieval can well capture this while the CTR model in image2product domain may easily be confused by the visual patterns. 
% CDAnet can transfer the knowledge of text2product domain into the image2product domain and improve the ranking results. 
In (b), we show a ranking example of a dress. Although our database does not have the the same item as the query, CDAnet has better ability of understanding which items are similar to the dress, while the online Base model ranked a skirt (circled with red box) in position 5. These cases help us to better understand the superiority of CDAnet.

% \usepackage{multirow}
\begin{table*}[]
\centering
\caption{Results to show that the translated features in book space have related item content as its original item content in movie space. The last column indicates whether the book title has close content as the movie.}
\vspace{-6pt}
% \vspace{-8pt}
\label{table:translated_analysis}
\renewcommand{\arraystretch}{1.0}
\setlength{\tabcolsep}{0.8mm}{ 
\scalebox{0.8}{
\begin{tabular}{c|ccl|cc|c}
\hline
                       & \multicolumn{3}{c|}{items in movie domain}                                                                                                             & \multicolumn{2}{c|}{items of 5-nearest neighbours from Book domain}   & \multirow{2}{*}{content match ?} \\ \cline{1-6}
UserID                 & \multicolumn{1}{c|}{ItemID}                 & \multicolumn{1}{c|}{Movie name}                                    & Movie genre                         & \multicolumn{1}{c|}{ItemID} & Book name                               &                                  \\ \hline
\multirow{5}{*}{21778} & \multicolumn{1}{c|}{\multirow{5}{*}{8329}}  & \multicolumn{1}{c|}{\multirow{5}{*}{The Addams Family}}            & \multirow{5}{*}{emotion, love}      & \multicolumn{1}{c|}{39040}  & Where My Heart Breaks                   & Yes                              \\ \cline{5-7} 
                       & \multicolumn{1}{c|}{}                       & \multicolumn{1}{c|}{}                                              &                                     & \multicolumn{1}{c|}{33024}  & The Program                             & No                               \\ \cline{5-7} 
                       & \multicolumn{1}{c|}{}                       & \multicolumn{1}{c|}{}                                              &                                     & \multicolumn{1}{c|}{29368}  & Lick (A Stage Dive Novel)               & Yes                              \\ \cline{5-7} 
                       & \multicolumn{1}{c|}{}                       & \multicolumn{1}{c|}{}                                              &                                     & \multicolumn{1}{c|}{37926}  & Snowed Over                             & Yes                              \\ \cline{5-7} 
                       & \multicolumn{1}{c|}{}                       & \multicolumn{1}{c|}{}                                              &                                     & \multicolumn{1}{c|}{44092}  & Where the Stars Still Shine             & Yes                              \\ \hline
\multirow{5}{*}{4842}  & \multicolumn{1}{c|}{\multirow{5}{*}{13887}} & \multicolumn{1}{c|}{\multirow{5}{*}{Mystery Science Theater 3000}} & \multirow{5}{*}{mystery, thrillers} & \multicolumn{1}{c|}{10505}  & A Wild Sheep Chase: A Novel             & Yes                              \\ \cline{5-7} 
                       & \multicolumn{1}{c|}{}                       & \multicolumn{1}{c|}{}                                              &                                     & \multicolumn{1}{c|}{7713}   & Wyrd Sisters                            & Yes                              \\ \cline{5-7} 
                       & \multicolumn{1}{c|}{}                       & \multicolumn{1}{c|}{}                                              &                                     & \multicolumn{1}{c|}{8205}   & Polar Star                              & Yes                              \\ \cline{5-7} 
                       & \multicolumn{1}{c|}{}                       & \multicolumn{1}{c|}{}                                              &                                     & \multicolumn{1}{c|}{135765} & Harry Potter and the Chamber of Secrets & Yes                              \\ \cline{5-7} 
                       & \multicolumn{1}{c|}{}                       & \multicolumn{1}{c|}{}                                              &                                     & \multicolumn{1}{c|}{6374}   & The Ambler Warning                      & Yes                              \\ \hline
\multirow{5}{*}{2078}  & \multicolumn{1}{c|}{\multirow{5}{*}{4977}}  & \multicolumn{1}{c|}{\multirow{5}{*}{Stargate SG-1 Season 2}}       & \multirow{5}{*}{fantasy, romance}   & \multicolumn{1}{c|}{43606}  & Countess So Shameless                   & No                               \\ \cline{5-7} 
                       & \multicolumn{1}{c|}{}                       & \multicolumn{1}{c|}{}                                              &                                     & \multicolumn{1}{c|}{16339}  & Crimson City                            & Yes                              \\ \cline{5-7} 
                       & \multicolumn{1}{c|}{}                       & \multicolumn{1}{c|}{}                                              &                                     & \multicolumn{1}{c|}{35940}  & Not Your Ordinary Wolf Girl             & Yes                              \\ \cline{5-7} 
                       & \multicolumn{1}{c|}{}                       & \multicolumn{1}{c|}{}                                              &                                     & \multicolumn{1}{c|}{10612}  & The Resisters                           & Yes                              \\ \cline{5-7} 
                       & \multicolumn{1}{c|}{}                       & \multicolumn{1}{c|}{}                                              &                                     & \multicolumn{1}{c|}{12671}  & Dagger-Star (Epic of Palins, Book 1)    & Yes                              \\ \hline
\end{tabular}
}}
\vspace{-6pt}
\end{table*}

\subsection{Content of Translated Features}
CDAnet learns how to translate the latent features between domains and exploits these translated features as augmentation information to boost the target domain CTR prediction, so it is curious to see whether the translated features are meaningful for CTR prediction. In this part, we design an experiment to see whether the latent features have related content before and after translation. 

To be specific on Amazon dataset, given a user $u$ that has training instances sharing positive labels in both domains\footnote{\textbf{CDAnet does not require a user to have training instances in both domains. We control the user variable here to better analyze the translated features.} We let the instances in two domains share the positive label here for better analysis, because the ``negative'' instances in CTR are usually sampled and have less confidence than positives.}, we can obtain the latent feature matrix $\boldsymbol{Z_{b}^u}\in \mathbb{R}^{N_{b}^u\times d}$ of book domain and $\boldsymbol{Z_{m}^u}\in \mathbb{R}^{N_{m}^u\times d}$ of movie domain, where $N_{b}^u$ and $N_{m}^u$ denote the number of positive training instances for user $u$ in book domain and movie domain respectively. 
Then, we can get the translated features of $\boldsymbol{Z_{m}^u}$ by the learned translator and it is denoted as $\boldsymbol{Z_{m}^{u'}}\in \mathbb{R}^{N_{m}^u\times d}$. 
Next, for a row (i.e. a sample contains user and item features) in the the translated feature $\boldsymbol{Z_{m}^{u'}}$, we find its $k$-nearest neighbours in $\boldsymbol{Z_{b}^u}$. In this case, we denote the book names from these neighbors as the content information of the translated feature. Finally, for a row in $\boldsymbol{Z_{m}^{u'}}$, we can check whether the content information of translated features in book space is related to the original movie name in movie space. The results are summarized in Table~\ref{table:translated_analysis}.

From this Table, we find that the 5-nearest neighbour books have close content to the corresponding movie. For example, given UserID ``21778'', the interacted the movie is ``The Addams Family'' which tells an emotional and love story. Meanwhile, for the corresponding translated features, the 5-nearest neighbor books are also about emotion and love. It shows that the translated features can capture related content from movie domain to book domain. 
% This result matches a common belief in CDCTR that if a user likes an item in one domain, the user may also love the items sharing related content in other domains~\cite{li2019ddtcdr,liu2022continual,chen2020towards}. 
Note CDAnet does not use any alignment information of items across domains, but it can automatically capture this and provide additional information to boost the CTR prediction in target domain. 
% This further verifies the value of CDAnet's translation network.

\begin{table}[]
\centering
\caption{Ablation study on different model parts of CDAnet. We take CDAnet+MMOE here as an example. ``w/o'' means without the corresponding module. ``w/o translation network'' means we directly train the augmentation network which is random initialized. ``w/o augmentation network'' means we directly use the translation network for evaluation.}
\vspace{-6pt}
% \vspace{-8pt}
\label{table:different_modules}
\renewcommand{\arraystretch}{0.8}
\setlength{\tabcolsep}{1.0mm}{ 
\scalebox{0.9}{
\begin{tabular}{ccccc}
\hline
Dataset                   & \multicolumn{2}{c}{Amazon} & \multicolumn{2}{c}{Taobao} \\ \hline
Domain                    & movie        & book        & ad           & rec         \\ \hline
w/o $\mathcal{L}^{orth}$ & 0.7153       & 0.7734      & 0.6190      & 0.7022      \\
w/o $\mathcal{L}^{cross}$ & 0.7188       & 0.7757      & 0.6191      & 0.7003      \\
w/o translation network     & 0.6464       & 0.7649      & 0.6075            & 0.7001           \\
w/o augmentation network    & 0.6818       & 0.7686      & 0.6168       & 0.6811      \\
CDAnet                     & \textbf{0.7225}       & \textbf{0.7811}      & \textbf{0.6200}      & \textbf{0.7034}      \\ \hline
\end{tabular}
}}
\vspace{-12pt}
\end{table}
% \subsection{Ablation Study}
\subsection{Impacts of Different Model Parts}
% To investigate the impacts of different model parts, we conduct an experiment to see the performance change when removing the corresponding module. 
To assess the consequences of varying model components, we undertook an experiment to analyze the effects on performance when removing the associated module. The results are shown in Table~\ref{table:different_modules}.
% Comparing the result between CDAnet and w/o MMOE, we see the MMOE-based feature extractor has positive effects on boosting model performance. This MMOE module may help translation network alleviate optimization conflict issue and better transfer knowledge. 

Considering the result of w/o $\mathcal{L}^{orth}$ and w/o $\mathcal{L}^{cross}$,
we see either removing the orthogonal constraint loss or removing the cross-supervision loss would cause deteriorated performance. The orthogonal constraint loss $\mathcal{L}^{orth}$ can help CDAnet keep the similarity among latent features after translation and avoid mode collapse problem. The cross-supervision loss $\mathcal{L}^{cross}$ plays a vital role in learning two translators. 
% Without $\mathcal{L}^{cross}$, the translators cannot learn how to translate latent features into another space. 
Further, either removing translation network or augmentation network would cause rather poor performance. The translation network learns how to transfer knowledge between domains. 
When removing it, we cannot have useful knowledge for later augmentation network. 
The augmentation network reuses the pre-trained parameters of translation network and employs the additional translated latent features for final CTR prediction. When removing it, we only have the translation network whose goal is translation and cannot guarantee good CTR prediction performance.

\section{Conclusion and Future Work}
Cross-domain click-through rate (CDCTR) prediction is an important research topic in real-world systems. Compared to most existing methods that rely on different implicit ways to transfer knowledge, we propose a novel model named CDAnet for explicit, \emph{flexible} and \emph{efficient} knowledge transfer. \emph{CDAnet contains a translation stage that translates the latent feature of target domain into source domain view and augmentation stage that uses this translated feature to augment the target domain model learning.} Through extensive experiments, we show CDAnet is able to learn meaningful translated features and boost the target domain CTR prediction performance. Results on the large-scale production dataset and online system at Taobao app show its superiority in real-world applications. 
Despite of this, CDAnet still has inadequacies in transferring knowledge. 
Some ingredients of the latent features may not be transferrable and have negative effects in translation. In the future, we will study a more effective technique for feature translation.
% Despite of this, CDAnet still has inadequacies in transferring knowledge. 
% The latent features of two domains are usually not fully overlapped, which means some ingredients of the latent features may not be translated and may have negative effects in translation.
% In the future, we will study a more effective technique for feature translation.

% \begin{acks}
% \end{acks}

\newpage
\bibliographystyle{ACM-Reference-Format}
\bibliography{sample-sigconf}


\begin{thebibliography}{38}

%%% the Web address.

\ifx \showCODEN    \undefined \def \showCODEN     #1{\unskip}     \fi
\ifx \showDOI      \undefined \def \showDOI       #1{#1}\fi
\ifx \showISBNx    \undefined \def \showISBNx     #1{\unskip}     \fi
\ifx \showISBNxiii \undefined \def \showISBNxiii  #1{\unskip}     \fi
\ifx \showISSN     \undefined \def \showISSN      #1{\unskip}     \fi
\ifx \showLCCN     \undefined \def \showLCCN      #1{\unskip}     \fi
\ifx \shownote     \undefined \def \shownote      #1{#1}          \fi
\ifx \showarticletitle \undefined \def \showarticletitle #1{#1}   \fi
\ifx \showURL      \undefined \def \showURL       {\relax}        \fi
% invisible to TeX
\providecommand\bibfield[2]{#2}
\providecommand\bibinfo[2]{#2}
\providecommand\natexlab[1]{#1}
\providecommand\showeprint[2][]{arXiv:#2}

\bibitem[Chen et~al\mbox{.}(2021)]%
        {chen2021user}
\bibfield{author}{\bibinfo{person}{Lei Chen}, \bibinfo{person}{Fajie Yuan}, \bibinfo{person}{Jiaxi Yang}, \bibinfo{person}{Xiangnan He}, \bibinfo{person}{Chengming Li}, {and} \bibinfo{person}{Min Yang}.} \bibinfo{year}{2021}\natexlab{}.
\newblock \showarticletitle{User-specific Adaptive Fine-tuning for Cross-domain Recommendations}.
\newblock \bibinfo{journal}{\emph{IEEE Transactions on Knowledge and Data Engineering}} (\bibinfo{year}{2021}).
\newblock


\bibitem[Chen et~al\mbox{.}(2020)]%
        {chen2020towards}
\bibfield{author}{\bibinfo{person}{Xu Chen}, \bibinfo{person}{Ya Zhang}, \bibinfo{person}{Ivor~W Tsang}, \bibinfo{person}{Yuangang Pan}, {and} \bibinfo{person}{Jingchao Su}.} \bibinfo{year}{2020}\natexlab{}.
\newblock \showarticletitle{Towards equivalent transformation of user preferences in cross domain recommendation}.
\newblock \bibinfo{journal}{\emph{ACM Transactions on Information Systems (TOIS)}} (\bibinfo{year}{2020}).
\newblock


\bibitem[Cheng et~al\mbox{.}(2016)]%
        {cheng2016wide}
\bibfield{author}{\bibinfo{person}{Heng-Tze Cheng}, \bibinfo{person}{Levent Koc}, \bibinfo{person}{Jeremiah Harmsen}, \bibinfo{person}{Tal Shaked}, \bibinfo{person}{Tushar Chandra}, \bibinfo{person}{Hrishi Aradhye}, \bibinfo{person}{Glen Anderson}, \bibinfo{person}{Greg Corrado}, \bibinfo{person}{Wei Chai}, \bibinfo{person}{Mustafa Ispir}, {et~al\mbox{.}}} \bibinfo{year}{2016}\natexlab{}.
\newblock \showarticletitle{Wide \& deep learning for recommender systems}. In \bibinfo{booktitle}{\emph{Proceedings of the 1st workshop on deep learning for recommender systems}}. \bibinfo{pages}{7--10}.
\newblock


\bibitem[Guo et~al\mbox{.}(2017)]%
        {10.5555/3172077.3172127}
\bibfield{author}{\bibinfo{person}{Huifeng Guo}, \bibinfo{person}{Ruiming Tang}, \bibinfo{person}{Yunming Ye}, \bibinfo{person}{Zhenguo Li}, {and} \bibinfo{person}{Xiuqiang He}.} \bibinfo{year}{2017}\natexlab{}.
\newblock \showarticletitle{DeepFM: A Factorization-Machine Based Neural Network for CTR Prediction}. In \bibinfo{booktitle}{\emph{Proceedings of the 26th International Joint Conference on Artificial Intelligence}} (Melbourne, Australia) \emph{(\bibinfo{series}{IJCAI'17})}. \bibinfo{publisher}{AAAI Press}, \bibinfo{pages}{1725–1731}.
\newblock
\showISBNx{9780999241103}


\bibitem[Hou et~al\mbox{.}(2023)]%
        {hou2023deep}
\bibfield{author}{\bibinfo{person}{Xuyang Hou}, \bibinfo{person}{Zhe Wang}, \bibinfo{person}{Qi Liu}, \bibinfo{person}{Tan Qu}, \bibinfo{person}{Jia Cheng}, {and} \bibinfo{person}{Jun Lei}.} \bibinfo{year}{2023}\natexlab{}.
\newblock \showarticletitle{Deep Context Interest Network for Click-Through Rate Prediction}.
\newblock \bibinfo{journal}{\emph{Conference on Information and Knowledge Management}} (\bibinfo{year}{2023}).
\newblock


\bibitem[Hu et~al\mbox{.}(2018)]%
        {hu2018conet}
\bibfield{author}{\bibinfo{person}{Guangneng Hu}, \bibinfo{person}{Yu Zhang}, {and} \bibinfo{person}{Qiang Yang}.} \bibinfo{year}{2018}\natexlab{}.
\newblock \showarticletitle{Conet: Collaborative cross networks for cross-domain recommendation}. In \bibinfo{booktitle}{\emph{International conference on information and knowledge management}}. ACM, \bibinfo{pages}{667--676}.
\newblock


\bibitem[Huan et~al\mbox{.}(2023)]%
        {10.1145/3580305.3599955}
\bibfield{author}{\bibinfo{person}{Zhaoxin Huan}, \bibinfo{person}{Ang Li}, \bibinfo{person}{Xiaolu Zhang}, \bibinfo{person}{Xu Min}, \bibinfo{person}{Jieyu Yang}, \bibinfo{person}{Yong He}, {and} \bibinfo{person}{Jun Zhou}.} \bibinfo{year}{2023}\natexlab{}.
\newblock \showarticletitle{SAMD: An Industrial Framework for Heterogeneous Multi-Scenario Recommendation}. In \bibinfo{booktitle}{\emph{Proceedings of the 29th ACM SIGKDD Conference on Knowledge Discovery and Data Mining}} (Long Beach, CA, USA) \emph{(\bibinfo{series}{KDD '23})}. \bibinfo{publisher}{Association for Computing Machinery}, \bibinfo{address}{New York, NY, USA}, \bibinfo{pages}{4175–4184}.
\newblock
\showISBNx{9798400701030}
\urldef\tempurl%
\url{https://doi.org/10.1145/3580305.3599955}
\showDOI{\tempurl}


\bibitem[Isola et~al\mbox{.}(2017)]%
        {isola2017image}
\bibfield{author}{\bibinfo{person}{Phillip Isola}, \bibinfo{person}{Jun-Yan Zhu}, \bibinfo{person}{Tinghui Zhou}, {and} \bibinfo{person}{Alexei~A Efros}.} \bibinfo{year}{2017}\natexlab{}.
\newblock \showarticletitle{Image-to-image translation with conditional adversarial networks}. In \bibinfo{booktitle}{\emph{Proceedings of the IEEE conference on computer vision and pattern recognition}}. \bibinfo{pages}{1125--1134}.
\newblock


\bibitem[Kim et~al\mbox{.}(2017)]%
        {kim2017learning}
\bibfield{author}{\bibinfo{person}{Taeksoo Kim}, \bibinfo{person}{Moonsu Cha}, \bibinfo{person}{Hyunsoo Kim}, \bibinfo{person}{Jung~Kwon Lee}, {and} \bibinfo{person}{Jiwon Kim}.} \bibinfo{year}{2017}\natexlab{}.
\newblock \showarticletitle{Learning to discover cross-domain relations with generative adversarial networks}. In \bibinfo{booktitle}{\emph{International conference on machine learning}}. PMLR, \bibinfo{pages}{1857--1865}.
\newblock


\bibitem[Li et~al\mbox{.}(2015)]%
        {li2015click}
\bibfield{author}{\bibinfo{person}{Cheng Li}, \bibinfo{person}{Yue Lu}, \bibinfo{person}{Qiaozhu Mei}, \bibinfo{person}{Dong Wang}, {and} \bibinfo{person}{Sandeep Pandey}.} \bibinfo{year}{2015}\natexlab{}.
\newblock \showarticletitle{Click-through prediction for advertising in twitter timeline}. In \bibinfo{booktitle}{\emph{Proceedings of the 21th ACM SIGKDD International Conference on Knowledge Discovery and Data Mining}}. \bibinfo{pages}{1959--1968}.
\newblock


\bibitem[Li and Tuzhilin(2020)]%
        {li2019ddtcdr}
\bibfield{author}{\bibinfo{person}{Pan Li} {and} \bibinfo{person}{Alexander Tuzhilin}.} \bibinfo{year}{2020}\natexlab{}.
\newblock \showarticletitle{DDTCDR: Deep Dual Transfer Cross Domain Recommendation}.
\newblock \bibinfo{journal}{\emph{International conference on web search and data mining}} (\bibinfo{year}{2020}).
\newblock


\bibitem[Liu et~al\mbox{.}(2022)]%
        {liu2022continual}
\bibfield{author}{\bibinfo{person}{Lixin Liu}, \bibinfo{person}{Yanling Wang}, \bibinfo{person}{Tianming Wang}, \bibinfo{person}{Dong Guan}, \bibinfo{person}{Jiawei Wu}, \bibinfo{person}{Jingxu Chen}, \bibinfo{person}{Rong Xiao}, \bibinfo{person}{Wenxiang Zhu}, {and} \bibinfo{person}{Fei Fang}.} \bibinfo{year}{2022}\natexlab{}.
\newblock \showarticletitle{Continual Transfer Learning for Cross-Domain Click-Through Rate Prediction at Taobao}.
\newblock \bibinfo{journal}{\emph{arXiv preprint arXiv:2208.05728}} (\bibinfo{year}{2022}).
\newblock


\bibitem[Liu et~al\mbox{.}(2017)]%
        {liu2017unsupervised}
\bibfield{author}{\bibinfo{person}{Ming-Yu Liu}, \bibinfo{person}{Thomas Breuel}, {and} \bibinfo{person}{Jan Kautz}.} \bibinfo{year}{2017}\natexlab{}.
\newblock \showarticletitle{Unsupervised image-to-image translation networks}.
\newblock \bibinfo{journal}{\emph{Advances in neural information processing systems}}  \bibinfo{volume}{30} (\bibinfo{year}{2017}).
\newblock


\bibitem[Ma et~al\mbox{.}(2018)]%
        {ma2018modeling}
\bibfield{author}{\bibinfo{person}{Jiaqi Ma}, \bibinfo{person}{Zhe Zhao}, \bibinfo{person}{Xinyang Yi}, \bibinfo{person}{Jilin Chen}, \bibinfo{person}{Lichan Hong}, {and} \bibinfo{person}{Ed~H Chi}.} \bibinfo{year}{2018}\natexlab{}.
\newblock \showarticletitle{Modeling task relationships in multi-task learning with multi-gate mixture-of-experts}. In \bibinfo{booktitle}{\emph{Proceedings of the 24th ACM SIGKDD international conference on knowledge discovery \& data mining}}. \bibinfo{pages}{1930--1939}.
\newblock


\bibitem[Masoudnia and Ebrahimpour(2014)]%
        {masoudnia2014mixture}
\bibfield{author}{\bibinfo{person}{Saeed Masoudnia} {and} \bibinfo{person}{Reza Ebrahimpour}.} \bibinfo{year}{2014}\natexlab{}.
\newblock \showarticletitle{Mixture of experts: a literature survey}.
\newblock \bibinfo{journal}{\emph{Artificial Intelligence Review}}  \bibinfo{volume}{42} (\bibinfo{year}{2014}), \bibinfo{pages}{275--293}.
\newblock


\bibitem[Ouyang et~al\mbox{.}(2019)]%
        {ouyang2019deep}
\bibfield{author}{\bibinfo{person}{Wentao Ouyang}, \bibinfo{person}{Xiuwu Zhang}, \bibinfo{person}{Li Li}, \bibinfo{person}{Heng Zou}, \bibinfo{person}{Xin Xing}, \bibinfo{person}{Zhaojie Liu}, {and} \bibinfo{person}{Yanlong Du}.} \bibinfo{year}{2019}\natexlab{}.
\newblock \showarticletitle{Deep spatio-temporal neural networks for click-through rate prediction}. In \bibinfo{booktitle}{\emph{Proceedings of the 25th ACM SIGKDD International Conference on Knowledge Discovery \& Data Mining}}. \bibinfo{pages}{2078--2086}.
\newblock


\bibitem[Ouyang et~al\mbox{.}(2020)]%
        {ouyang2020minet}
\bibfield{author}{\bibinfo{person}{Wentao Ouyang}, \bibinfo{person}{Xiuwu Zhang}, \bibinfo{person}{Lei Zhao}, \bibinfo{person}{Jinmei Luo}, \bibinfo{person}{Yu Zhang}, \bibinfo{person}{Heng Zou}, \bibinfo{person}{Zhaojie Liu}, {and} \bibinfo{person}{Yanlong Du}.} \bibinfo{year}{2020}\natexlab{}.
\newblock \showarticletitle{Minet: Mixed interest network for cross-domain click-through rate prediction}. In \bibinfo{booktitle}{\emph{Proceedings of the 29th ACM international conference on information \& knowledge management}}. \bibinfo{pages}{2669--2676}.
\newblock


\bibitem[Pang et~al\mbox{.}(2021)]%
        {pang2021image}
\bibfield{author}{\bibinfo{person}{Yingxue Pang}, \bibinfo{person}{Jianxin Lin}, \bibinfo{person}{Tao Qin}, {and} \bibinfo{person}{Zhibo Chen}.} \bibinfo{year}{2021}\natexlab{}.
\newblock \showarticletitle{Image-to-image translation: Methods and applications}.
\newblock \bibinfo{journal}{\emph{IEEE Transactions on Multimedia}} (\bibinfo{year}{2021}).
\newblock


\bibitem[Rendle(2010)]%
        {rendle2010factorization}
\bibfield{author}{\bibinfo{person}{Steffen Rendle}.} \bibinfo{year}{2010}\natexlab{}.
\newblock \showarticletitle{Factorization machines}. In \bibinfo{booktitle}{\emph{2010 IEEE International conference on data mining}}. IEEE, \bibinfo{pages}{995--1000}.
\newblock


\bibitem[Richardson et~al\mbox{.}(2007)]%
        {richardson2007predicting}
\bibfield{author}{\bibinfo{person}{Matthew Richardson}, \bibinfo{person}{Ewa Dominowska}, {and} \bibinfo{person}{Robert Ragno}.} \bibinfo{year}{2007}\natexlab{}.
\newblock \showarticletitle{Predicting clicks: estimating the click-through rate for new ads}. In \bibinfo{booktitle}{\emph{Proceedings of the 16th international conference on World Wide Web}}. \bibinfo{pages}{521--530}.
\newblock


\bibitem[Sener and Koltun(2018)]%
        {sener2018multi}
\bibfield{author}{\bibinfo{person}{Ozan Sener} {and} \bibinfo{person}{Vladlen Koltun}.} \bibinfo{year}{2018}\natexlab{}.
\newblock \showarticletitle{Multi-task learning as multi-objective optimization}.
\newblock \bibinfo{journal}{\emph{Advances in neural information processing systems}}  \bibinfo{volume}{31} (\bibinfo{year}{2018}).
\newblock


\bibitem[Sheng et~al\mbox{.}(2021)]%
        {sheng2021one}
\bibfield{author}{\bibinfo{person}{Xiang-Rong Sheng}, \bibinfo{person}{Liqin Zhao}, \bibinfo{person}{Guorui Zhou}, \bibinfo{person}{Xinyao Ding}, \bibinfo{person}{Binding Dai}, \bibinfo{person}{Qiang Luo}, \bibinfo{person}{Siran Yang}, \bibinfo{person}{Jingshan Lv}, \bibinfo{person}{Chi Zhang}, \bibinfo{person}{Hongbo Deng}, {et~al\mbox{.}}} \bibinfo{year}{2021}\natexlab{}.
\newblock \showarticletitle{One model to serve all: Star topology adaptive recommender for multi-domain ctr prediction}. In \bibinfo{booktitle}{\emph{Proceedings of the 30th ACM International Conference on Information \& Knowledge Management}}. \bibinfo{pages}{4104--4113}.
\newblock


\bibitem[Tang et~al\mbox{.}(2020)]%
        {tang2020progressive}
\bibfield{author}{\bibinfo{person}{Hongyan Tang}, \bibinfo{person}{Junning Liu}, \bibinfo{person}{Ming Zhao}, {and} \bibinfo{person}{Xudong Gong}.} \bibinfo{year}{2020}\natexlab{}.
\newblock \showarticletitle{Progressive layered extraction (ple): A novel multi-task learning (mtl) model for personalized recommendations}. In \bibinfo{booktitle}{\emph{Fourteenth ACM Conference on Recommender Systems}}. \bibinfo{pages}{269--278}.
\newblock


\bibitem[Vandenhende et~al\mbox{.}(2021)]%
        {vandenhende2021multi}
\bibfield{author}{\bibinfo{person}{Simon Vandenhende}, \bibinfo{person}{Stamatios Georgoulis}, \bibinfo{person}{Wouter Van~Gansbeke}, \bibinfo{person}{Marc Proesmans}, \bibinfo{person}{Dengxin Dai}, {and} \bibinfo{person}{Luc Van~Gool}.} \bibinfo{year}{2021}\natexlab{}.
\newblock \showarticletitle{Multi-task learning for dense prediction tasks: A survey}.
\newblock \bibinfo{journal}{\emph{IEEE transactions on pattern analysis and machine intelligence}} (\bibinfo{year}{2021}).
\newblock


\bibitem[Wang et~al\mbox{.}(2022)]%
        {wang2022enhancing}
\bibfield{author}{\bibinfo{person}{Fangye Wang}, \bibinfo{person}{Yingxu Wang}, \bibinfo{person}{Dongsheng Li}, \bibinfo{person}{Hansu Gu}, \bibinfo{person}{Tun Lu}, \bibinfo{person}{Peng Zhang}, {and} \bibinfo{person}{Ning Gu}.} \bibinfo{year}{2022}\natexlab{}.
\newblock \showarticletitle{Enhancing CTR prediction with context-aware feature representation learning}. In \bibinfo{booktitle}{\emph{Proceedings of the 45th International ACM SIGIR Conference on Research and Development in Information Retrieval}}. \bibinfo{pages}{343--352}.
\newblock


\bibitem[Wang et~al\mbox{.}(2017)]%
        {wang2017deep}
\bibfield{author}{\bibinfo{person}{Ruoxi Wang}, \bibinfo{person}{Bin Fu}, \bibinfo{person}{Gang Fu}, {and} \bibinfo{person}{Mingliang Wang}.} \bibinfo{year}{2017}\natexlab{}.
\newblock \showarticletitle{Deep \& cross network for ad click predictions}.
\newblock In \bibinfo{booktitle}{\emph{Proceedings of the ADKDD'17}}. \bibinfo{pages}{1--7}.
\newblock


\bibitem[Wang et~al\mbox{.}(2021)]%
        {wang2021dcn}
\bibfield{author}{\bibinfo{person}{Ruoxi Wang}, \bibinfo{person}{Rakesh Shivanna}, \bibinfo{person}{Derek Cheng}, \bibinfo{person}{Sagar Jain}, \bibinfo{person}{Dong Lin}, \bibinfo{person}{Lichan Hong}, {and} \bibinfo{person}{Ed Chi}.} \bibinfo{year}{2021}\natexlab{}.
\newblock \showarticletitle{Dcn v2: Improved deep \& cross network and practical lessons for web-scale learning to rank systems}. In \bibinfo{booktitle}{\emph{Proceedings of the Web Conference 2021}}. \bibinfo{pages}{1785--1797}.
\newblock


\bibitem[Wang et~al\mbox{.}(2020)]%
        {wang2020gradient}
\bibfield{author}{\bibinfo{person}{Zirui Wang}, \bibinfo{person}{Yulia Tsvetkov}, \bibinfo{person}{Orhan Firat}, {and} \bibinfo{person}{Yuan Cao}.} \bibinfo{year}{2020}\natexlab{}.
\newblock \showarticletitle{Gradient vaccine: Investigating and improving multi-task optimization in massively multilingual models}.
\newblock \bibinfo{journal}{\emph{arXiv preprint arXiv:2010.05874}} (\bibinfo{year}{2020}).
\newblock


\bibitem[Yang et~al\mbox{.}(2022)]%
        {yang2022click}
\bibfield{author}{\bibinfo{person}{Xiangli Yang}, \bibinfo{person}{Qing Liu}, \bibinfo{person}{Rong Su}, \bibinfo{person}{Ruiming Tang}, \bibinfo{person}{Zhirong Liu}, \bibinfo{person}{Xiuqiang He}, {and} \bibinfo{person}{Jianxi Yang}.} \bibinfo{year}{2022}\natexlab{}.
\newblock \showarticletitle{Click-through rate prediction using transfer learning with fine-tuned parameters}.
\newblock \bibinfo{journal}{\emph{Information Sciences}}  \bibinfo{volume}{612} (\bibinfo{year}{2022}), \bibinfo{pages}{188--200}.
\newblock


\bibitem[Yi et~al\mbox{.}(2017)]%
        {yi2017dualgan}
\bibfield{author}{\bibinfo{person}{Zili Yi}, \bibinfo{person}{Hao Zhang}, \bibinfo{person}{Ping Tan}, {and} \bibinfo{person}{Minglun Gong}.} \bibinfo{year}{2017}\natexlab{}.
\newblock \showarticletitle{Dualgan: Unsupervised dual learning for image-to-image translation}. In \bibinfo{booktitle}{\emph{Proceedings of the IEEE international conference on computer vision}}. \bibinfo{pages}{2849--2857}.
\newblock


\bibitem[Yu et~al\mbox{.}(2020)]%
        {yu2020gradient}
\bibfield{author}{\bibinfo{person}{Tianhe Yu}, \bibinfo{person}{Saurabh Kumar}, \bibinfo{person}{Abhishek Gupta}, \bibinfo{person}{Sergey Levine}, \bibinfo{person}{Karol Hausman}, {and} \bibinfo{person}{Chelsea Finn}.} \bibinfo{year}{2020}\natexlab{}.
\newblock \showarticletitle{Gradient surgery for multi-task learning}.
\newblock \bibinfo{journal}{\emph{Advances in Neural Information Processing Systems}}  \bibinfo{volume}{33} (\bibinfo{year}{2020}), \bibinfo{pages}{5824--5836}.
\newblock


\bibitem[Yuan et~al\mbox{.}(2019)]%
        {yuan2019darec}
\bibfield{author}{\bibinfo{person}{Feng Yuan}, \bibinfo{person}{Lina Yao}, {and} \bibinfo{person}{Boualem Benatallah}.} \bibinfo{year}{2019}\natexlab{}.
\newblock \showarticletitle{DARec: Deep Domain Adaptation for Cross-Domain Recommendation via Transferring Rating Patterns}.
\newblock \bibinfo{journal}{\emph{International joint conference on artificial intelligence}} (\bibinfo{year}{2019}).
\newblock


\bibitem[Yuksel et~al\mbox{.}(2012)]%
        {yuksel2012twenty}
\bibfield{author}{\bibinfo{person}{Seniha~Esen Yuksel}, \bibinfo{person}{Joseph~N Wilson}, {and} \bibinfo{person}{Paul~D Gader}.} \bibinfo{year}{2012}\natexlab{}.
\newblock \showarticletitle{Twenty years of mixture of experts}.
\newblock \bibinfo{journal}{\emph{IEEE transactions on neural networks and learning systems}} \bibinfo{volume}{23}, \bibinfo{number}{8} (\bibinfo{year}{2012}), \bibinfo{pages}{1177--1193}.
\newblock


\bibitem[Zhang et~al\mbox{.}(2023)]%
        {10.1145/3580305.3599758}
\bibfield{author}{\bibinfo{person}{Wei Zhang}, \bibinfo{person}{Pengye Zhang}, \bibinfo{person}{Bo Zhang}, \bibinfo{person}{Xingxing Wang}, {and} \bibinfo{person}{Dong Wang}.} \bibinfo{year}{2023}\natexlab{}.
\newblock \showarticletitle{A Collaborative Transfer Learning Framework for Cross-Domain Recommendation}. In \bibinfo{booktitle}{\emph{Proceedings of the 29th ACM SIGKDD Conference on Knowledge Discovery and Data Mining}} (Long Beach, CA, USA) \emph{(\bibinfo{series}{KDD '23})}. \bibinfo{publisher}{Association for Computing Machinery}, \bibinfo{address}{New York, NY, USA}, \bibinfo{pages}{5576–5585}.
\newblock
\showISBNx{9798400701030}
\urldef\tempurl%
\url{https://doi.org/10.1145/3580305.3599758}
\showDOI{\tempurl}


\bibitem[Zhang et~al\mbox{.}(2022)]%
        {zhang2022keep}
\bibfield{author}{\bibinfo{person}{Yujing Zhang}, \bibinfo{person}{Zhangming Chan}, \bibinfo{person}{Shuhao Xu}, \bibinfo{person}{Weijie Bian}, \bibinfo{person}{Shuguang Han}, \bibinfo{person}{Hongbo Deng}, {and} \bibinfo{person}{Bo Zheng}.} \bibinfo{year}{2022}\natexlab{}.
\newblock \showarticletitle{KEEP: An Industrial Pre-Training Framework for Online Recommendation via Knowledge Extraction and Plugging}. In \bibinfo{booktitle}{\emph{Proceedings of the 31st ACM International Conference on Information \& Knowledge Management}}. \bibinfo{pages}{3684--3693}.
\newblock


\bibitem[Zhou et~al\mbox{.}(2019)]%
        {zhou2019deep}
\bibfield{author}{\bibinfo{person}{Guorui Zhou}, \bibinfo{person}{Na Mou}, \bibinfo{person}{Ying Fan}, \bibinfo{person}{Qi Pi}, \bibinfo{person}{Weijie Bian}, \bibinfo{person}{Chang Zhou}, \bibinfo{person}{Xiaoqiang Zhu}, {and} \bibinfo{person}{Kun Gai}.} \bibinfo{year}{2019}\natexlab{}.
\newblock \showarticletitle{Deep interest evolution network for click-through rate prediction}. In \bibinfo{booktitle}{\emph{Proceedings of the AAAI conference on artificial intelligence}}, Vol.~\bibinfo{volume}{33}. \bibinfo{pages}{5941--5948}.
\newblock


\bibitem[Zhou et~al\mbox{.}(2018)]%
        {zhou2018deep}
\bibfield{author}{\bibinfo{person}{Guorui Zhou}, \bibinfo{person}{Xiaoqiang Zhu}, \bibinfo{person}{Chenru Song}, \bibinfo{person}{Ying Fan}, \bibinfo{person}{Han Zhu}, \bibinfo{person}{Xiao Ma}, \bibinfo{person}{Yanghui Yan}, \bibinfo{person}{Junqi Jin}, \bibinfo{person}{Han Li}, {and} \bibinfo{person}{Kun Gai}.} \bibinfo{year}{2018}\natexlab{}.
\newblock \showarticletitle{Deep interest network for click-through rate prediction}. In \bibinfo{booktitle}{\emph{Proceedings of the 24th ACM SIGKDD international conference on knowledge discovery \& data mining}}. \bibinfo{pages}{1059--1068}.
\newblock


\bibitem[Zhu et~al\mbox{.}(2017)]%
        {zhu2017unpaired}
\bibfield{author}{\bibinfo{person}{Jun-Yan Zhu}, \bibinfo{person}{Taesung Park}, \bibinfo{person}{Phillip Isola}, {and} \bibinfo{person}{Alexei~A Efros}.} \bibinfo{year}{2017}\natexlab{}.
\newblock \showarticletitle{Unpaired image-to-image translation using cycle-consistent adversarial networks}. In \bibinfo{booktitle}{\emph{Proceedings of the IEEE international conference on computer vision}}. \bibinfo{pages}{2223--2232}.
\newblock


\end{thebibliography}

\appendix

\begin{figure}[h]
\centering
\begin{minipage}[t]{0.23\textwidth}
\centering
\includegraphics[width=\textwidth]{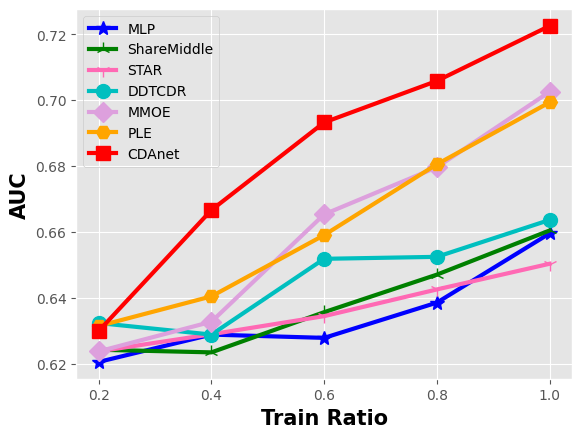}
\vspace{-20pt}
\caption*{(a) \footnotesize{Amazon-movie}}
\end{minipage}
\begin{minipage}[t]{0.23\textwidth}
\centering
\includegraphics[width=\textwidth]{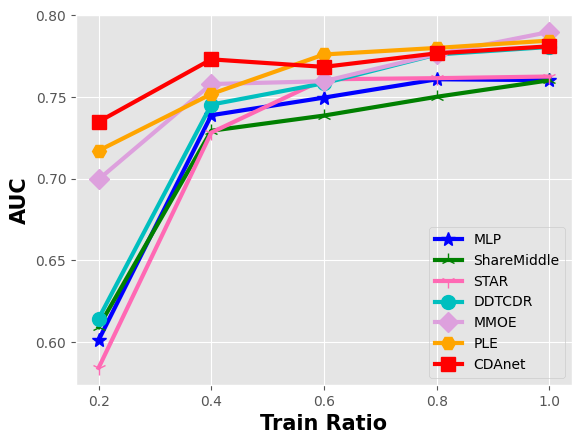}
\vspace{-20pt}
\caption*{(b) \footnotesize{Amazon-book}}
\end{minipage}\\
\begin{minipage}[t]{0.23\textwidth}
\centering
\includegraphics[width=\textwidth]{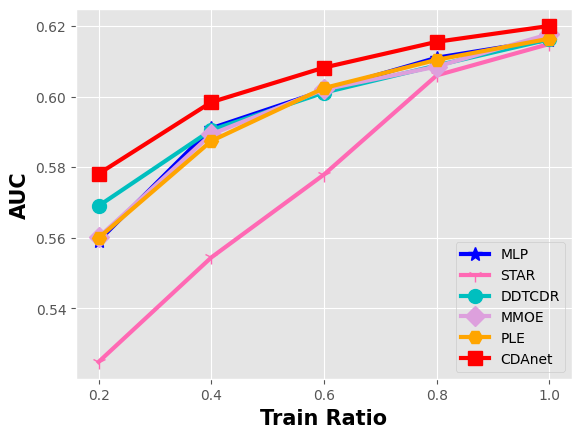}
\vspace{-20pt}
\caption*{(c) \footnotesize{Taobao-ad}}
\end{minipage}
\begin{minipage}[t]{0.23\textwidth}
\centering
\includegraphics[width=\textwidth]{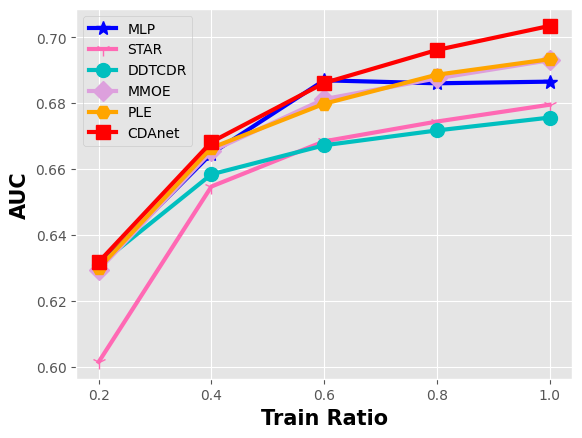}
\vspace{-20pt}
\caption*{(d) \footnotesize{Taobao-rec}}
\end{minipage}
\vspace{-6pt}
\caption{The effects of different sparsity levels on Amazon and Taobao. Train ratio means the ratio of the original train data. We take CDAnet+MMOE here as an example.}
\label{figure:sparse_data}
% \vspace{-10pt}
\end{figure}

\begin{figure}[h]
\centering
\begin{minipage}[t]{0.23\textwidth}
\centering
\includegraphics[width=\textwidth]{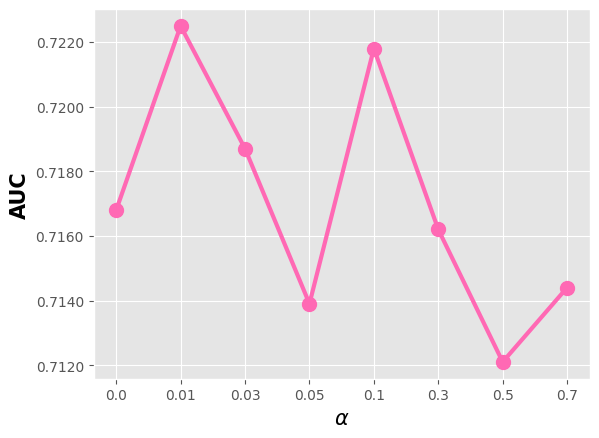}
\vspace{-20pt}
\caption*{(a) \footnotesize{$\alpha$ on Amazon-movie}}
\end{minipage}
\begin{minipage}[t]{0.23\textwidth}
\centering
\includegraphics[width=\textwidth]{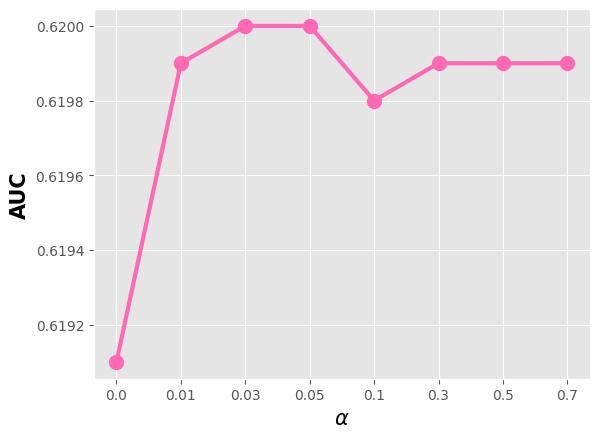}
\vspace{-20pt}
\caption*{(b) \footnotesize{$\alpha$ on Taobao-ad}}
\end{minipage}\\
\begin{minipage}[t]{0.23\textwidth}
\centering
\includegraphics[width=\textwidth]{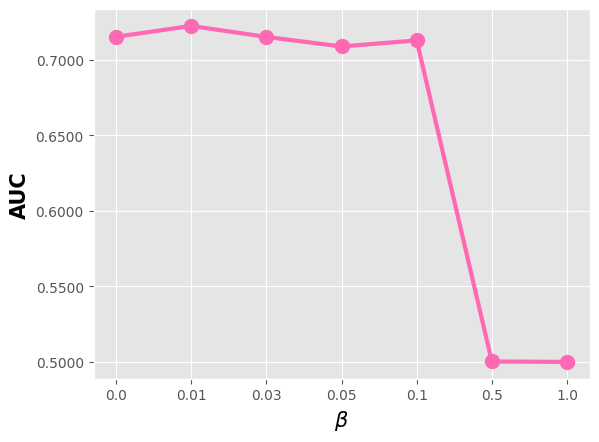}
\vspace{-20pt}
\caption*{(c) \footnotesize{$\beta$ on Amazon-movie}}
\end{minipage}
\begin{minipage}[t]{0.23\textwidth}
\centering
\includegraphics[width=\textwidth]{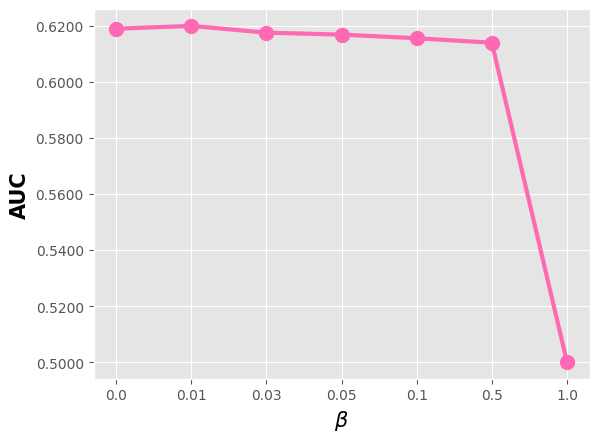}
\vspace{-20pt}
\caption*{(d) \footnotesize{$\beta$ on Taobao-ad}}
\end{minipage}
\vspace{-6pt}
\caption{The effects of hyper-parameters on different datasets. $\alpha$ and $\beta$ are loss weights on different loss parts. We take CDAnet+MMOE here as an example.}
% \vspace{-10pt}
\label{figure:different_hyperparameters}
\end{figure}
\section{Ablation Study}
\subsection{Different Sparse Data}
In order to investigate how CDAnet performs under more sparse conditions, we conduct an experiment to see the model performance at different sparsity levels of the training data.
In particular, fixing the validation and test set, we vary the ratio of the original train data to the new train set. 
The results are shown in Figure~\ref{figure:sparse_data}, where we can see that CDAnet can achieve better results at almost all sparsity levels compared to other cross-domain CTR methods. CDAnet has a better way to transfer knowledge, so that it can show more generalized performance.

\subsection{Parameter Sensitivity}
In CDAnet, $\alpha$ and $\beta$ control the loss weight on cross-supervision loss and orthogonal constraint loss, respectively. Here, we study the parameter sensitivity of these hyper-parameters to investigate their effects on model performance. The results are shown in Figure~\ref{figure:different_hyperparameters}. 

In this figure, $\alpha$ controls the strength of cross-supervision on learning the translators. Too large $\alpha$ may dominate the learning of $\mathcal{L}^{vani}$ and cause bad effects on learning cross-supervised translators. $\beta$ controls the weight of $\mathcal{L}^{orth}$ and too large $\beta$ may limit the learning flexibility of the translators. These two hyper-parameters could be selected according to the validation set performance on different datasets.

\end{document}